\documentclass[12pt]{article}

\usepackage{graphicx}
\usepackage{amsmath,amssymb}
\usepackage{bm}
\allowdisplaybreaks
\begin{document}
\baselineskip=0.7cm
\renewcommand{\figurename}{Fig.\@}
\renewcommand{\thesection}{\arabic{section}.}
\renewcommand{\theequation}{\arabic{section}.\arabic{equation}}
\renewcommand{\thesubsection}{\arabic{section}.\arabic{subsection}}
\makeatletter
\def\section{\@startsection{section}{1}{\z@}{-3.5ex plus -1ex minus 
 -.2ex}{2.3ex plus .2ex}{\large}} 
\def\subsection{\@startsection{subsection}{2}{\z@}{-3.25ex plus -1ex minus 
 -.2ex}{1.5ex plus .2ex}{\normalsize\it}}
 \def\subsubsection{\@startsection{subsubsection}{3}{\z@}{-3.25ex plus -1ex minus 
 -.2ex}{1.5ex plus .2ex}{\normalsize\it}}

\makeatother
\makeatletter
\def\lesim{\mathrel{\mathpalette\gl@align<}}
\def\gtsim{\mathrel{\mathpalette\gl@align>}}
\def\gl@align#1#2{\lower.7ex\vbox{\baselineskip\z@skip\lineskip.2ex%
  \ialign{$\m@th#1\hfil##\hfil$\crcr#2\crcr\sim\crcr}}}
\makeatother

\makeatletter
\def\@cite#1#2{$^{\hbox{\scriptsize{#1\if@tempswa , #2\fi})}}$}
\makeatletter

\def\thefootnote{\alph{footnote}}


\newcommand{\sn}{{\rm\,sn\,}}
\newcommand{\cn}{{\rm\,cn\,}}
\newcommand{\dn}{{\rm\,dn\,}}

\vspace*{0.5cm}
\begin{center}
\Large 
Statistical Field Theory \\of Interacting Nambu Dynamics
\vspace{0.3cm}

\normalsize
 \vspace{0.4cm}
Tamiaki {\sc  Yoneya} \footnote{
Emeritus Professor}

\vspace{0.3cm}

{\it Institute of Physics, University of Tokyo\\
Komaba, Meguro-ku, Tokyo 153-8902}

\vspace{1cm}
Abstract
\end{center}
We develop a statistical field theory for classical Nambu dynamics 
by employing partially the method of quantum field theory. 
One of unsolved problems in Nambu dynamics has been 
to extend it to interacting systems without violating a 
generalized canonical structure associated with the presence of multiple Hamiltonians, which together govern the dynamics of time evolution on 
an equal footing. 
In the present paper, we propose to include 
interactions from the standpoint of classical statistical dynamics 
by formulating it as a field theory 
on Nambu's generalized phase space in an operator formalism. 
We first construct a general framework for such a field theory 
and its probabilistic interpretation. Then, on the basis 
of this new framework, 
we give a simple model of self-interactions in a 
many-body Nambu system treated as a closed dynamical system 
satisfying the H-theorem. It is shown that a generalized micro-canonical ensemble and a generalized canonical ensemble  
characterized by many temperatures  
are reached dynamically as equilibrium states 
starting with certain classes of initial non-equilibrium states via continuous 
Markov processes. Compared with the usual classical 
statistical mechanics on the basis 
of standard Hamiltonian dynamics, some important new features 
associated with Nambu dynamics will 
emerge, with respect to the symmetries underlying dynamics of the 
non-equilibrium as well as the equilibrium states and also to some 
conceptual properties, such as a formulation of a generalized 
KMS-like 
condition characterizing the generalized canonical equilibrium states and a 
`relative' nature of the temperatures. 

\newpage
\section{Introduction}
\subsection{Backgrounds and motivations}
Nambu's generalized Hamilton equations of motion\cite{nambu1} 
in the simplest case takes the following form\footnote{Unless otherwise 
 stated explicitly, the Einstein convention 
 with respect to the coordinate indices is assumed 
 throughout the present paper. The metric is flat, 
$g_{ij}=g^{ij}=\delta_{ij}$, and we freely use both 
upper and lower indices, $x^i=x_i$ for convenience of 
expressing equations. Also we 
use abbreviations such as $\partial_i=\partial/\partial x^i$. } 
\begin{align}
\frac{d x^i}{dt}=\epsilon^{ijk}\partial_jH\partial_kK
=\{H, K, x^i\}\equiv X^i, 
\label{nambueq}
\end{align}
where $H=H(x^i)$ and $K=K(x^i)$ are two independent 
functions of the phase-space coordinates $(x^1,x^2,x^3)$ 
in three dimensions. The bracket symbol, called the Nambu bracket, 
in this expression is 
defined in terms of the Jacobian corresponding to 
transformation of $(x^1,x^2,x^3)$ to a 
set of arbitrary three functions 
$(A,B,C)$.
\begin{align}
	\{A,B,C\}\equiv \frac{\partial(A,B,C)}{\partial(x^1,x^2,x^3)}=\epsilon^{ijk}
	\partial_iA\partial_jB\partial_kC.
\end{align}
Then, obviously, $H$ and $K$ are both conserved, 
\begin{align}
	\frac{dH}{dt}=\dot{x}^i\partial_iH=\epsilon^{ijk}\partial_jH\partial_kK \partial_iH=0, 
	\quad 	\frac{dK}{dt}=\dot{x}^i\partial_iK=\epsilon^{ijk}\partial_jH\partial_kK \partial_iK=0. 
	\nonumber
\end{align}
It is also clear that the system satisfies 
the Liouville equation
\begin{align}
\partial_iX^i=0, 
\end{align}
which guarantees that the volume of phase space 
occupied by an aggregate of systems described 
by the same equations of motion is conserved. 

Thus, instead of a single Hamiltonian 
in the ordinary Hamiltonian dynamics for a 
conventional even-dimensional phase space, Nambu's generalized 
Hamiltonian dynamics is governed by two Hamiltonian-like 
conserved quantities $H$ and $K$ which give the time 
evolution of the system together on an equal footing 
in the three-dimensional phase space $(x^1,x^2,x^3)$. 
One of the main motivations behind his 
proposal was to construct a generalized 
statistical mechanics such 
that a canonical ensemble is 
characterized by a weight factor with two temperature-like parameters corresponding to 
a generalized Boltzmann distribution in phase space,	$$	e^{-\beta H-\gamma K}.	$$

He emphasized possible physical relevance of 
this generalization by pointing out that 
the Euler equations of motion for a 
rigid rotator 
can be cast in the above form by 
identifying $H$ and $K$ as
\begin{align}
	H=\frac{1}{2}(L_1^2+L_2^2+L_3^2), \qquad K=\frac{1}{2}\bigg(\frac{L_1^2}{I_1}+
	\frac{L_2^2}{I_2}+\frac{L_3^2}{I_3}\bigg),
	\label{Eulertop}
\end{align}
and choosing the phase-space coordinates to be the 
components themselves of angular-momentum vector in the body-fixed 
frame, $(x^1,x^2,x^3)=(L_1,L_2,L_3)$. 
This formalism can naturally be generalized to 
$n$ dimensional phase space $(x^1,x^2,\ldots,x^n)$ 
with the equations of motion for any $n\ge 2$,
\begin{align}
	\dot{x}^i&=\epsilon^{ij_1\cdots j_{n-1}}
	\partial_{j_1}H_1\partial_{j_2}H_2
	\cdots \partial_{j_{n-1}}H_n
	=\frac{\partial(H_1,H_2,\ldots,H_{n-1},x^i)}{\partial
	(x^1,x^2,\ldots,x^{n})}.\nonumber\\
	&\equiv \{H_1,H_2,\ldots,H_{n-1},x^i\}.
\end{align}
Now there exists a set of $n-1$ conserved Hamiltonian-like 
functions, 
$(H_1,H_2,\ldots,H_{n-1})$, which 
are independent to each other. 

Nambu mainly devoted 
himself to a general discussion of canonical transformations 
and possibilities toward quantization. 
However, he encountered certain 
obstacles which hindered straightforward generalizations of the 
structure intrinsic to the conventional Hamiltonian formalism. 
In particular, by extending the idea to 
$3N$ dimensions with $N$ 
triplets $(x^1_a,x^2_a,x^3_a)$ $(a=1,\ldots,N)$ in analogy with 
ordinary Hamiltonian dynamics,  
he found that linear 
canonical transformations mixing different $a$ indices 
did not work as he desired. Namely, only transformations 
within each single triplet work appropriately. 
Perhaps for this reason, no further discussion about the statistical aspect of 
the problem has been attempted, and a majority of later efforts following his 
proposal have been aiming 
toward the problem of quantization.

With this situation in mind, the purpose 
of the present paper is to initiate a 
discussion about the statistical aspect of 
Nambu dynamics, and to 
lay a foundation along this direction 
proposing a basic framework toward 
generalized statistical mechanics in 
a form of statistical field theory.  
In developing such a framework anew, we will take a standpoint that the Nambu 
dynamics is not a mere 
amendment to the conventional Hamiltonian dynamics: rather, 
we take the view that his proposal 
amounts to positing a hypothetical world governed by a new extended  
dynamical framework. 

In general, when we consider a system with a large number of 
constituents, such as a gas consisting of $N$ molecules, 
it is convenient to consider a collection of 
systems of the identical structure, namely an {\it ensemble} of systems, and treat them by a statistical method. 
Usually, the so-called microcanonical 
or canonical ensemble, depending on the situations,  
is {\it postulated} as a basis for describing equilibrium 
states on the basis of probabilistic arguments.
However, from a {\it physical} standpoint of 
{\it dynamics}, 
the microstate of each member system 
in the ensemble 
evolves following the equations of motion autonomously, and the distribution 
of possible states of systems in the ensemble is 
determined by the distribution of initial conditions which are in principle 
completely arbitrary for each member system 
independently of the other systems in the ensemble. 
Thus from the viewpoint of dynamics, a crucial basic 
question concerning the realization of equilibrium states is whether there is any natural dynamical 
mechanism such that the distribution of 
states evolves into a definite equilibrium distribution 
despite the complete arbitrariness of initial condition in each system. 
From this point of view, it is not an easy task to 
justify the probabilistic arguments. 

Now, we have to emphasize that, compared with the situation of the 
conventional Hamilton dynamics, 
Nambu dynamics
is quite problematical at  
a foundational level. 
Namely, Nambu dynamics has been known to be rigid in a rather 
stringent way: 
with the proviso that the equations of motion are 
invariant under general canonical transformations, 
it cannot be extended suitably 
to {\it interacting} systems 
by generalizing the original $n$-dimensional phase space 
$(x^1,x^2,\ldots, x^{n})\in \mathbb{R}^{n}$ to 
$Nn$-dimensional multiple phase spaces 
$x=(x^1_a,x^2_a,\ldots,x^n_a) \,\, (a=1,2,\ldots,N) \in \mathbb{R}^{Nn}$ 
equipped with the canonical structure
\[	\{x_a^{i_1},x_a^{i_2},\ldots, x_a^{i_n}\}=\epsilon^{i_1
i_2\cdots i_n}
\]
for each $a$ with all other cases of 
mixed $a$ indices vanishing, {\it except} for the case 
where the system is completely separated 
into $N$ independent systems without any 
interaction among them. In other words, 
there is a difficulty in generalizing the conserved 
energies $H_k(x)$ $(k=1,\ldots,n-1)$ 
in such a way that they explicitly involve `potential' 
energies corresponding to the existence of interactions  
among different systems. Indeed, this was essentially the hindrance  
Nambu faced with as alluded to above. 
It is clear that, if there were {\it no interaction} of any 
kind among 
different degrees of freedom discriminated by the $a$-indices above, the distribution 
of $n-1$ energies $H_k(x)$ in the ensemble could never change from the 
initial one, and therefore that there is {\it no} reason {\it at all}  
to expect the emergence of any equilibrium 
distribution {\it dynamically}. Furthermore, with no 
interaction whatsoever, familiar thermodynamical 
concepts such as, say, heat bath and thermal contact of 
different systems 
obeying Nambu dynamics would be {\it wholly groundless}. 

Thus in developing the statistical mechanics of Nambu dynamics logically from a dynamical perspective, 
it seems imperative for us to begin its construction from scratch and 
devise some novel approaches for overcoming the above 
difficulties related to interactions.
In this paper, we shall assume as a {\it working hypothesis} 
that arbitrary two systems in the ensemble can exchange their  
energies $H_k$ instantaneously between them, such that the 
sums of the energies is strictly preserved for each $k$, but the 
initial conditions of both 
systems are updated at every such instant of energy 
exchange. 
Nothing in principle prevents us from assuming 
such an axiom for constructing a new framework, as long as the invariance 
of the system under canonical transformations of 
the phases space coordinates, which is one of 
the key features of Nambu dynamics, is maintained throughout 
this process. 
Except for these instantaneous interactions, 
the system evolves deterministically obeying the 
Nambu equations of motion. Here it is important to 
recognize that,  
due to the first-order nature of the 
Nambu equations of motion, `collision' 
which is local in a literal sense with respect to 
phase-space coordinates does {\it not} make sense at all. Namely, 
such interactions are feasible {\it only 
if} they are non-local to a certain extent with respect 
to the phase coordinates $x_a^i$. 
Suppose that such non-local updates of initial 
conditions are repeated innumerable times 
for all possible pairs 
of systems in the ensemble without any preference. 
Then, the distribution 
of states may change as a genuine dynamical process, due to an `ergodic' 
(or `chaotic') 
mixing 
of initial 
conditions, and 
the system is expected to reach a statistical equilibrium after a sufficient 
lapse of time. 
We shall propose a simple model of such dynamical processes
by formulating a {\it statistical} 
field theory 
defined on the Nambu phase space, which can 
fittingly be utilized for a concrete 
realization of the above intuitive picture of 
non-local interactions. From a formalistic viewpoint, our approach will follow closely 
the methods of quantum field theory, despite the fact that  
we are treating classical Nambu systems, since 
the fundamental dynamical degrees of freedom are supposed to be 
field operators defined appropriately on the base space which is 
nothing but the phase space-time itself.

\subsection{Contents of the present paper}
The next section, consisted of 4 subsections, provides preliminaries to the present paper for the purpose of making the present paper 
reasonably self-contained: we 
start with the Liouville equation
for the 
Nambu equations of motion without interactions. 
However, in order to incorporate the 
non-local interactions mentioned above, we have to break through such a mild framework, 
since non-locality of self-interactions necessarily forces us 
to go beyond the realm of continuous dynamical processes 
on which the Liouville equation is based. The situation is in stark   contrast to 
that in standard statistical mechanics based on the 
conventional Hamiltonian formalism, where we are 
always allowed to 
assume the validity of the Liouville equation with an 
appropriate Hamiltonian including interaction potentials. 
In developing such a new framework, we will stress an important 
symmetry, called the 
`$\mathcal{N}$-symmetry', which characterizes in crucial ways the 
structure of Nambu dynamics equipped with 
two (or more) Hamiltonian-like functions, as discussed 
emphatically in \cite{yoneintro}.

After finishing the preparations as above, 
two main sections, sections 3 and 4, are devoted to constructing a framework of our whole discussions: 
in section 3 (consisted of 4 subsections),  
we shall first introduce the field operators which create 
and annihilates a single Nambu system as 
basic dynamical variables defined on the 
Nambu phase space-time. 
Then we develop an operator formalism 
for {\it classical} statistical mechanics, aimed toward 
interacting many-body Nambu systems later. 
 In section 4 (consisted of 2 subsections), we construct a general framework of 
classical probabilistic interpretation for statistical 
states on which the field operators are acting. 
Then in the next two main sections we shall 
devote ourselves to proposing and studying a specific model of the non-local interaction, 
and examining its consequences in detail: in section 5 (consisted of 6 subsections), we postulate a fundamental dynamical equation, 
called the `master equation',  
which governs the approach of the systems to 
equilibrium states as a continuous Markov process. 
It will be established that the evolutions  
described by the master equation  
satisfies in general the `H-theorem'. Then, we shall propose a simple 
solvable model for the non-local 
interaction and analyze its properties in detail. 
In the final section 6 (consisted of 4 subsections), the equilibrium states will be 
derived on the 
basis of the H-theorem, 
placing emphasis on its nature which 
arises owing to the presence of two (or more) 
Hamiltonian-like functions, in particular, 
related to the $\mathcal{N}$-symmetry. Then we will give 
a new characterization 
of the equilibrium states by extending a classical version of 
the so-called KMS condition, which is familiar in the standard quantum 
statistical mechanics, to a two-component vector form 
such that it is covariant under the $\mathcal{N}$-symmetry 
transformations. 

Although in the main text we treat only the case $n=3$ explicitly 
for the purpose of making our arguments as concrete as possible in a 
simplest nontrivial setting, 
it must be fairly obvious to serious readers 
that the whole of our discussions is extended to general $n$-dimensional cases straightforwardly. 

There are two short appendices: in appendix A, for the purpose of making a 
comparison with the approach adopted in the main text, we will 
briefly discuss a more phenomenological approach to statistical 
equilibrium states for Nambu dynamics, a Fokker-Planck-type formalism 
in $n$ dimensions which has been familiar in general statistical physics, originated in Einstein's theory of Brownian motion 
more than a century ago. 
In appendix B, we present a simplified matrix-model analog  
which captures some crucial aspects of the kernel function, introduced  in section 5 for constructing a concrete model for 
the non-local interaction. 

\section{Preliminaries}
\setcounter{equation}{0}
\subsection{Nambu equations of motion in terms of the Liouville equation}

A basis for starting our discussions of statistical mechanics of 
Nambu dynamics is Liouville's theorem.
The Liouville equation of Nambu 
dynamics for the density function $\rho(x,t)$ 
in Nambu phase space, which is treated as a 
3-dimensional Euclidean space with the Descartes 
coordinates $x\equiv (x^1,x^2,x^3)$, is
\begin{align}
\label{liouville}
	\frac{\partial \rho}{\partial t}+X^i(x)\partial_i\rho=0,
\end{align}
with
	$X^i(x)\equiv \epsilon^{ijk}\partial_jH(x)\partial_kK(x)$. 
The connection of this equation with 
the Nambu equations of motion can be made manifest by considering 
the Green function for \eqref{liouville}:
\begin{align}
	\bigg(\frac{\partial}{\partial t}+X^i(x)\frac{\partial}{\partial x^i}\bigg)
	G_{\mathrm{r}}(x,t;x_0,t_0)=\delta^3(x-x_0)\delta(t-t_0),\nonumber
\end{align}
under the initial condition 
\begin{align}
	\lim_{t\rightarrow t_0+}G_{\mathrm{r}}(x,t;x_0,t_0)=\delta^3(x-x_0).\nonumber
\end{align}
The solution is uniquely determined for $t>t_0$ to be
\begin{align}
	G_{\mathrm{r}}(x,t;x_0,t_0)=\delta^3\big(x-x(t-t_0;x_0)\big),
	\nonumber
\end{align}
where $x^i(t-t_0;x_0)$ is the solution of the 
Nambu equations of motion \eqref{nambueq} with the initial 
condition $x^i(0;x_0)=x_0^i$: in fact, we have 
\begin{align}
	\partial_t\delta^3\big(x-x(t-t_0;x_0)\big)
&=-\dot{x}^i(t;x_0,t_0)\partial_i\delta^3\big(x-
x(t-t_0;x_0)\big)\nonumber\\
&=-X^i\big(x(t-t_0;x_0)\big)\partial_i\delta^3\big(x-
x(t;x_0,t_0)\big)\nonumber \\
&=-X^i(x)\partial_i\delta^3\big(x-
x(t-t_0;x_0)\big),\nonumber
\end{align}
where, in the last equality, use has been made of 
a crucial property that the flows described by the Nambu equations 
of motion satisfy the condition of 
incompressibility, or equivalently, Liouville's theorem, 
\begin{align}
\label{liouvilltheorem}
	\partial_iX^i(x)=0.
\end{align}


\subsection{The Liouville equation as a Hamilton-Jacobi theory 
of Nambu dynamics} 
From a {\it purely formal} standpoint, the Liouville equation 
\eqref{liouville} can also be regarded, by multiplying Planck constant 
and the associated imaginary unit which are cancelled out automatically, 
as a (time-dependent) Schr\"{o}dinger equation with $\pi_i\equiv -i\hbar\partial_i$ 
being the canonical momentum operator that is 
conjugate to $x^i$. Note that 
in this interpretation, the phase space coordinate $x^i$ 
is now treated as the canonical coordinates in 
six dimensional generalized phase space $(x^i,\pi_i)$. 
From this viewpoint, a peculiarity 
arising from the linearity with respect to $\pi_i$ 
of the corresponding `Hamiltonian' 
\begin{align}
\label{hamil}
	H_0\equiv \pi_iX^i(x)
\end{align}
 is that there is no direct relation 
between velocity vector $\dot{x}^i$ and 
momentum vector $\pi_i$, and consequently the former is 
diagonalized simultaneously as coordinate vector $x^i$. 
In other words, because of this linearity, 
its Hamilton-Jacobi equation 
is essentially equivalent 
to Schr\"{o}dinger equation. This conforms to the `classicality' 
of our formalism: there is no spreading of wave packet without any contradiction with 
the uncertainty principle for the {\it enlarged} phase space 
$(x^i,\pi_i)$.
If we extend the system by adding, for example, the usual kinetic term 
as
\begin{align}
	H_0\rightarrow H_0'=\pi_iX^i+\frac{\pi_i\pi^i}{2m},\nonumber
\end{align}
the wave packets would necessarily have a quantum mechanical 
spreading, corresponding to the equations of motion 
$\dot{x}^i=X^i+\pi^i/m$ which demands 
the uncertainty with respect to velocity associated 
with the uncertainty of $\pi^i/m$: the Nambu system can be 
regarded as the limit of infinite mass, $m\rightarrow\infty$, and the 
absence of dispersion (or the spreading of a wave packet) is simply 
a consequence of this special limit. This formal 
analogy provides a rationale for 
introducing field operators as fundamental degrees of 
freedom in representing genuinely 
classical many-body systems in the next section. 

The specific form, \eqref{hamil}, of our Hamiltonian 
can be characterized by a `gauge' symmetry $\delta_{\lambda}H_0=0$ 
with\footnote{Actually, the naming `gauge' here is a 
misnomer, based only on a superficial formal analogy, because $\pi_i$ 
itself is not a gauge field 
in any sense. However, for later convenience, we use 
this convention.}
\begin{align}
\label{gaugesym}
	\delta_{\lambda} \pi_i=\partial_i\lambda (H,K)
	=\partial_H\lambda\partial_iH+\partial_K\lambda\partial_iK,
\end{align}
where $\lambda=\lambda(H,K)$ is an arbitrary function 
of two variables $(H,K)$, 
due essentially to the condition 
$\partial_iHX^i=0=\partial_iKX^i$. 
The latter property immediately yields the conservation laws,
$
\dot{\lambda}(H,K)=0, 
$
as a special case of Noether's theorem. 
We call the quantities $H(x),K(x)$ `energy functions' in what follows, being the analogs of energy 
in ordinary classical mechanics, 
although there is an important difference 
from the ordinary energies of conventional Hamiltonian dynamics: 
 even if the energy functions $H,K$ have stationary points only at 
isolated points as in the case of ordinary energies, the stationary 
(or fixed) points of the 
trajectories satisfying the Nambu equations of motion 
in general form continuous one-dimensional curves,
since $X^i=0$ whenever two gradient vectors 
$\nabla H$ and $\nabla K$ are orthogonal to each other.
Thus, a peculiarity of Nambu dynamics with respect to Noether's theorem, 
at least in the present formalism, is that the Hamiltonian \eqref{hamil} itself associated with 
time-translation symmetry does {\it not} have the meaning and role of 
energy.  In connection with this, it is also to be 
noticed that $H_0$, being equal to 
a difference of two positive quantities 
$H_0=\frac{1}{4}\big[(\pi_i+X^i)^2-(\pi_i-X^i)^2\big]$,  
has no bound at all, and hence cannot play the 
role of energy: at best it could be 
used as a certain constraint for allowed states, 
in analogy with, say, the Hamiltonian constraint 
in the Hamiltonian formulation of gravity. 
For more details of the Hamilton-Jacobi theory of Nambu dynamics, the readers are referred to \cite{yoneintro} where the variational principle of Nambu dynamics is also treated from a coherent and unified standpoint.

\subsection{The $\mathcal{N}$-symmetry}
As for the symmetry of the Hamiltonian $H_0$, it is important to notice that $X^i$ itself is invariant under 
the transformation 
$(H,K)\rightarrow (H',K')$ of energy functions such 
that 
\begin{align}
	\frac{\partial(H',K')}{\partial(H,K)}=1, 
	\label{Ngauge}
\end{align}
which leads to 
\begin{align}
	\{H',K',x^i\}=\frac{1}{2}\epsilon^{ijk}
	\frac{\partial(H',K')}{\partial (H,K)}
	\frac{\partial(H,K)}{\partial(x^j,x^k)}=
	\{H,K,x^i\}.\nonumber
\end{align}
In terms of differential forms, we can 
express this invariance as
\begin{align}
	dH\wedge dK-dH'\wedge dK'=d(HdK-H'dK')=0.
	\label{diffinv}
\end{align}
Obviously, any pair $(H,K)$ obtained 
by this transformation plays completely 
equivalent role in the dynamics of time evolution 
of the phase coordinates $x^i$.  
For later convenience, let 
us call this symmetry the `$\mathcal{N}$-symmetry'. Since 
the transformation $(H,K)\rightarrow (H'.K')$ is, 
as a consequence of \eqref{diffinv}, generated by a single 
arbitrary function $F=F(K,K')$ defined by
\begin{align}
	HdK-H'dK'=dF 
	\quad \Leftrightarrow \quad
	\frac{\partial F}{\partial K}=H, \quad 
	\frac{\partial F}{\partial K'}=-H', \nonumber
\end{align}
the number of the true degrees of freedom of the energy functions driving time evolution as exhibited 
in the specific form of the Nambu equations motion 
is in fact one, in conformity with the existence 
of the single Hamiltonian $H_0$ which 
is invariant under the $\mathcal{N}$-transformations, which is 
formally a sort of `gauge' transformation when $HdK$ is 
interpreted as the Clebsch representation for a gauge 
potential in the space of all possible energy functions.

%
\subsection{Free many-body systems}
For a generic system with free $N$-body Nambu systems with the 
coordinates $(x_1,\ldots, x_N)$, 
the density function $\rho(x,t)$ can  naturally be 
expressed (for $t>t_0$) as a formal superposition 
of that in a single-body case,
\begin{align}
\label{densityN}
	&\rho(x,t)=\sum_{a=1}^NG_{\mathrm{r}}(x,t:x_{a},t_0)=
	\sum_{a=1}^N\delta^3\big(x-x_a(t-t_0;x_{a})\big), \quad \int \rho(x,t)d^3x =N,
\end{align}
where the initial condition for the system $a$ 
is $x^i_a(t_0)=x^i_{a}$ $(a=1,\ldots,N)$, 
which still satisfies the same Liouville equation \eqref{liouville}, 
owing to the linearity of the latter.
In terms of the density function, the conservation law for 
the energy functions $(H,K)$, for instance, 
is generalized to $N$-body systems: 
\begin{align}
	\frac{d}{dt}\int \lambda(H(x),K(x))\rho(x,t)d^3x
	&=-\int \lambda(H(x),K(x))X^i(x)\partial_i\rho(x,t)d^3x 
	\nonumber\\
	&=\int \partial_i\lambda(H(x),K(x))X^i(x)\rho(x,t)d^3x
	=0. \label{consrho}
\end{align}
Note that 
\begin{align}
	\int \lambda(H(x),K(x))\rho(x,t)d^3x=\sum_{a=1}^n
	\lambda\big(H(x_a),K(x_a)\big).\nonumber
\end{align}
The simplest case $\lambda=1$ gives the 
conservation of the number $N$ of Nambu particles 
in the system.

\section{A Field-Theory Formalism of Many-Body Nambu Systems}
\subsection{Field operators for Nambu particles}
Despite we veritably treat a classical system, in order to facilitate our statistical treatment of many-body systems 
with non-local interactions, we
 introduce a field operator $\psi(x,t)$ and 
its canonical conjugate $\psi^{\dagger}(x,t)$ as {\it basic 
dynamical degrees} of freedom, which we propose to call `Liouville fields' 
for convenience. They annihilates or 
creates, respectively,
 a Nambu system, which from now on we call a `Nambu particle',  at a phase-space point $x^i$ 
at time $t$. 
The corresponding vacuum (ket and bra) states are denoted by $|0\rangle$ 
and $\langle 0|$:
\begin{align}
	\label{fcommu}
	[\psi(x,t),\psi^{\dagger}(y,t)]&=\delta^3(x-y), 
	\quad [\psi(x,t),\psi(y,t)]=0=[\psi^{\dagger}(x,t),\psi^{\dagger}(y,t)],\\
	&\psi(x,t)|0\rangle=0=\langle 0|\psi^{\dagger}(x,t).
\end{align}
Throughout the present paper, we use bra-ket notations for classical 
many-body states on which the Liouville fields are operating.

The field equations for the Liouville fields are postulated to be  
\begin{align}
\label{ffeq}
\big(\partial_t+X^i(x)\partial_i\big)\psi(x,t)=0=
	\big(\partial_t+X^i(x)\partial_i\big)\psi^{\dagger}(x,t). 
\end{align}
We can easily check that the compatibility of the field equations 
with the commutation relations: 
for instance, taking time derivative of the first 
of \eqref{fcommu}, we find
\begin{align}
	&\partial_t[\psi(x,t),\psi^{\dagger}(y,t)]=
-\big(X^i(x)\partial_{x^i}+X^i(y)\partial_{y^i}\big)[\psi(x,t),\psi^{\dagger}(y,t)]\nonumber\\
&=-\big(X^i(x)-X^i(y)\big)\partial_{x^i}
\delta^3(x-y)=0
\nonumber\end{align}
using $\partial_iX^i=0$. 
The density 
function $\rho(x,t)$ corresponds to the 
operator $\psi^{\dagger}(x,t)\psi(x,t)$, so that
the Liouville equation \eqref{liouville} should 
now be regarded as a consequence of the field equation \eqref{ffeq}, 
justifying the naming `Liouville' fields. 
Their coincidence is due to the first-order nature, with respect to the 
derivatives,  
of these equations.

The Liouville field operators $\psi, \psi^{\dagger}$ enable us to reinterpret the single-body 
Green function $G_{\mathrm{r}}$ as the vacuum expectation 
value of a product of them placed at 
different space-time points, just as in quantum field theory: 
\begin{align}
\label{commupsipsidagger}
	G_{\mathrm{r}}(x,t:x_0,t_0)=\langle 0|T\big(\psi(x,t)\psi^{\dagger}(x_0,t_0)\big)|0\rangle
	=[\psi(x,t),\psi^{\dagger}(x_0,t_0)]
	\theta(t-t_0)
\end{align}
where $T$ is the usual time-ordering operator 
and the trivial factor $\langle 0|0\rangle\equiv 1$ 
is suppressed in the last equality. 
From the discussion given in the previous section for the 
Green function,  
this describes 
the {\it uniquely} determined trajectory of a single Nambu particle  
in terms of the field operators, with initial coordinates that are 
set to be $x_0^i$ at $t=t_0$, without any spreading.
Therefore our field-theory formalism in terms of the Liouville fields 
is completely equivalent to the usual approach 
of directly treating the individual equations of motion. 

It is important to notice that the field-theory 
formalism keeps the invariance of Nambu systems 
under canonical coordinate transformations: this is so 
in essentially the same sense as in the conventional 
Hamiltonian dynamics. 
The Liouville field operators as well as 
the energy functions are scalar, $\psi(x,t)=\psi'(x',t), H(x)=H'(x'), K(x)=K'(x')$, under general spatial coordinate 
transformation $x^i\rightarrow x'^i$. The defining properties of the Liouville fields 
and the equations satisfied by them are all form-invariant 
under the (time-independent) general canonical transformation 
which is nothing but the {\it volume-preserving} diffeomorphism,  
{\it vdiff}$_3$, 
 $x^i\rightarrow x'^i=x'^i(x)$, satisfying
\begin{align}
	\frac{\partial(x'^1,x'^2,x'^3)}{\partial(x^1,x^2,x^3)}=1. 
	\nonumber
\end{align} 
For example, the field equations take the following form 
in terms of Nambu bracket, 
\begin{align}
	&\partial_t\psi(x)=-\{H(x),K(x),\psi(x)\}=-\frac{\partial\big(H(x),K(x),\psi(x)\big)}{\partial(x^1,x^2,x^3)}, \nonumber
	\\
	&\partial_t\psi^{\dagger}(x)=-\{H(x),K(x),\psi^{\dagger}(x)\}=-\frac{\partial\big(H(x),K(x),\psi^{\dagger}(x)\big)}{\partial(x^1,x^2,x^3)},
	\nonumber
\end{align}
which are manifestly invariant under the canonical 
transformations. The whole formalism (including interactions to be included later) 
throughout the 
present paper will be invariant under the canonical transformations  
in the above sense.

The field equations in operator form are given 
by the Heisenberg-type equations of motion, 
\begin{align}
	\partial_t\psi(x,t)=[\hat{{\cal H}}_0,\psi(x,t)], 
	\quad \partial_t\psi^{\dagger}(x,t)=[\hat{{\cal H}}_0,
	\psi^{\dagger}(x,t)],\nonumber
\end{align}
by using the Hamiltonian operator
\begin{align}
	\hat{{\cal H}}_0&\equiv \int \psi^{\dagger}(x,t)H_0\psi(x,t)d^3x 
	\nonumber\\
	&=\int \psi^{\dagger}(x,t)X^i(x)
	\partial_i\psi(x,t)d^3x
	=-\int \partial_i\psi^{\dagger}(x,t)X^i(x)
	\psi(x,t)d^3x=-{\cal H}_0^{\dagger},\nonumber
\end{align}
which is by definition independent of time $t$ owing to the 
field equations and thus is 
consistent with $[\hat{{\cal H}}_0,\hat{{\cal H}}_0]\equiv 0$. 
Note the complete absence of imaginary unit 
in the present formalism.
For example, the 
time evolution with no wave-packet spreading is implemented by similarity 
transformations (in fact unitatry 
transformations due to anti-hermiticity of $\hat{{\cal H}}_0$) consistently  
as 
\begin{align}
	\psi(x,t)=e^{\hat{{\cal H}}_0(t-t_0)}\psi(x,t_0)
	e^{-\hat{{\cal H}}_0(t-t_0)}, 
	\quad \psi^{\dagger}(x,t)=e^{\hat{{\cal H}}_0(t-t_0)}\psi^{\dagger}(x,t_0)
	e^{-\hat{{\cal H}}_0(t-t_0)}.\nonumber
\end{align}

\subsection{Symmetry transformations and conservation laws 
in terms of field operators}

The conservation laws reflecting the gauge symmetry 
\eqref{gaugesym} are 
\begin{align}
\label{HGconsv}
	[\hat{{\cal H}}_0, \hat{H}]=0=[\hat{{\cal H}}_0, \hat{K}],
\end{align}
where we defined the energy operators,
\begin{align}
\label{HKhat}
	\hat{H}=\int \psi^{\dagger}(x,t)H(x)\psi(x,t)d^3x, 
	\quad \hat{K}=\int \psi^{\dagger}(x,t)K(x)\psi(x,t)d^3x,
\end{align}
which play the role of infinitesimal generators  
for the gauge transformation. 
We note that, 
so long as we consider only free $N$-body systems, the 
conservation laws of $H$ and $K$ are valid for each 
independent system separately: due to this, we can in fact 
generalize the conservation laws to 
\begin{align}
\label{conslaw}
	[\hat{{\cal H}}_0, \int \psi^{\dagger}
	(x,t)\lambda\big(H(x),K(x)\big)\psi(x,t)d^3x]=0
\end{align}
with arbitrary function $\lambda(H,K)$ as in \eqref{consrho}. 
The gauge transformations for the field operators 
generated by
\begin{align}
	\hat{\Lambda}(H,K)\equiv \int \psi^{\dagger}
	(x,t)\lambda\big(H(x),K(x)\big)\psi(x,t)d^3x\nonumber
\end{align}
 are 
\begin{align}
\begin{aligned}
\label{fgaugetrans}
	&\psi(x,t)\rightarrow e^{i\hat{\Lambda}(H,K)} \psi(x,t)e^{-i\hat{\Lambda}(H,k)}=e^{-i\lambda(H(x),K(x))}\psi(x,t), 
	\\ 
	&\psi^{\dagger}(x,t)\rightarrow e^{i\hat{\Lambda}(H,K)} \psi^{\dagger}(x,t)e^{-i\hat{\Lambda}(H,k)}=e^{i\lambda(H(x),K(x))}\psi^{\dagger}(x,t),
	\end{aligned}
\end{align}
under which the field equations and the Hamiltonian $\hat{\mathcal{H}}_0$ are invariant. 

However, as soon as non-local interactions that 
shuffle the initial conditions spontaneously 
are introduced as we have outlined in the previous section, 
the `gauge' invariance will necessarily be reduced a subgroup 
of all such transformations, namely, 
to the linear versions \eqref{HGconsv} of 
general conservation law 
 with a linear function
$\lambda(x,y)=c_1x+c_2y+c_0$ with constant $c_i$ ($i=0,1,2$). 
Of course, this does not cause any troublle at all, since there is 
no physical gauge field associated with this symmetry. 
For the same reason, the $\mathcal{N}$-symmetry \eqref{Ngauge} will also be 
reduced to its linearized form with constant (and real) matrix elements,
\begin{align}
	\begin{pmatrix}
		H' \\ K'
	\end{pmatrix}=\begin{pmatrix}
		a & b \\ c & d
	\end{pmatrix}
	\begin{pmatrix}
		H \\ K
	\end{pmatrix}, 
	\quad ad-bc=1,
	\label{sl2r}
\end{align}
constituting a group SL(2,$\mathbb{R}$).  
We have the same transformation laws for the 
operators $(\hat{H},\hat{K})$. 
Even in this reduced linearized form, the existence of such 
symmetry transformations of energy functions is an important key 
feature of 
Nambu dynamics, signifying its characteristic feature that has {\it no} 
analog in the conventional Hamiltonian dynamics. 
Keeping this symmetry 
as far as possible will be one of our guiding principles 
for developing an interacting field theory of Nambu dynamics. 
For later purpose of 
defining generalized canonical distributions, we assume that the energy functions $H$ and $K$ can be chosen 
to be non-negative and be increasing indefinitely 
for large absolute values of the 
phase coordinates, $|x|\rightarrow \infty$, by utilizing the 
$\mathcal{N}$-symmetry 
appropriately.

\subsection{Basis states for classical many-body Nambu systems}
Now, as a set of basis ket and bra-vectors for generic $n$-body states at time $t$, 
we can choose the following Fock states,
\begin{align}
\begin{aligned}
&|[x],t\rangle_{N}\equiv \frac{1}{\sqrt{N!}}\Psi^{\dagger}_N([x],t)|0\rangle, 
\quad \Psi^{\dagger}_N([x],t)
\equiv \prod_{a=1}^N\psi^{\dagger}(x_a,t).\\
&_N\langle [x],t|\equiv \frac{1}{\sqrt{N!}}\langle 0|\Psi_N([x],t), 
\quad \Psi_N([x],t)
\equiv \prod_{a=1}^N\psi(x_a,t),\label{nbody}
\end{aligned}
\end{align}
which obey the normalization condition
\begin{align}
\label{normcond}
	_N\langle [x],t|[x'],t\rangle_{N'}=\delta_{NN'}
	\frac{1}{N!}\sum_{{\rm P(a)}}
	\prod_{a=1}^N \delta^3(x_a-x'_{P(a)})
\end{align}
with $P(a)$ signifying the permutations of the 
indices $\{a\}=(1,2,\ldots, N)$. 
In this notation, the completeness relation of our classical phase space 
for arbitrary number of Nambu particles is expressed as
\begin{align}
	1=\sum_{N=0}^{\infty}
	\int 
	|[x],t)\rangle_N
	 \,_N\langle[x],t|[dx]_n, \label{comprelation}, \qquad [dx]_N\equiv \prod_{a=N}^n d^3x_a.
	\end{align}
Since the $N$-body composite operator 
defined here 
satisfies 
\begin{align}
\label{nbodycomp}
	\bigg(\frac{\partial}{\partial t}+\sum_{a=1}^NX^i(x_a)
	\frac{\partial}{\partial x_a^i}\bigg)\Psi^{\dagger}_N([x],t)=0, 
\end{align}
it is easy to check that the r.h.s of the completeness relation \eqref{comprelation} 
is independent of time $t$, by taking 
time derivative (and using partial integration) directly. 
The formal use of partial integration for operators is justified since 
the completeness relation is practically supposed to be used always 
for well-defined matrix elements.   
Note also that 
 the $N$-body operators $\Psi^{\dagger}_N([x],t)$ and $\Psi_N([x],t)$
 are totally symmetric under permutations 
of the coordinates because of the commutation relations \eqref{fcommu}, the same symmetry of the 
classical density function $\rho(x,t)$, \eqref{densityN}. 
Namely, we treat $N$ Nambu particles as 
indistinguishable, following the old 
proposal by Gibbs in his formulation of the principles of classical statistical 
mechanics. 
However, except for this indistinguishability, 
the above $N$-body state is still a {\it precise} (or 
dispersion-free) classical microstate with 
all the information available on a many-body system of Nambu particles at an instant of 
time $t$. 

In fact, the density operator defined by
\begin{align}
	\hat{\rho}(x,t)=\psi^{\dagger}(x,t)\psi(x,t),\nonumber
\end{align}
satisfies 
\begin{align}
\label{densi}
	\hat{\rho}(x,t)|[x],t_0\rangle_{N}
	=\rho(x,t)|[x],t_0\rangle_{N}, 
	\quad 
	\rho(x,t)=\sum_{a=1}^N\delta^3\big(x-x_a(t-t_0; x_a)\big), 
\end{align}
as is easily verified by a direct computation using \eqref{commupsipsidagger}. 
Namely, the basis state $|[x],t_0\rangle_N$ is an {\it eigenstate} 
of $\hat{\rho}(x,t)$ with the 
density function \eqref{densityN} 
being the corresponding eigenvalue.   
Conversely, it seems appropriate to say that the density 
operator characterizes classical states with precise and maximum information on many-body states of Nambu particles 
by the fact that such microstates are eigenstates 
of the density operator. 

However, even classically such an eigenstate with 
completely definite values for the coordinates 
of particles is a highly idealized concept: 
this is so for {\it any finite} $N\ge 2$, especially when particles are 
interacting with each other. 
It would be more natural and general to 
treat states with certain {\it statistical} spreading 
for each particle by introducing statistical 
ensembles.

\section{Statistical Ensembles  
in terms of Classical State Vectors}
\setcounter{equation}{0}
\subsection{Classical probability interpretation}
Let us now proceed to describe a statistical ensemble consisting of independent many-body systems, each with a fixed finite number ($N \ge 2$) of Nambu particles, in terms of our operator formalism. 
If we use the language of traditional classical statistical 
mechanics, the Liouville field operators $\psi(x,t)$ and $\psi^{\dagger}(x,t)$ 
are defined on the `$\mu-$space' of three dimensional 
Nambu phase space, while their action connects between the 
`$\varGamma$- spaces' of $3N$ dimensions with different $N$'s, 
either $N\rightarrow N-1$ or $N\rightarrow N+1$, respectively. 
For notational brevity, we suppress the index $N$ for the 
state vectors throughout 
the present section, since we consider statistical states 
with a fixed number ($N$) of Nambu particles.

We first note that  the microstates  
with definite coordinate values satisfy the following integral condition:
\begin{align}
	\int\, \langle [x],t|[x'],t\rangle
	[dx]=1=\int\, \langle [x],t|[x'],t\rangle
	[dx'].\nonumber
\end{align} 
Given an arbitrary, {\it real non-negative} function $f[x]=f(x_1,x_2,\ldots,x_N)$ ($\ge 0$), which is
  {\it totally symmetric} under arbitrary permutations of $N$ coordinates, 
we define a classical statistical state by
\begin{align}
\label{genesta}
	|f(t_0)\rangle\equiv \int 
	f[x]|[x],t_0\rangle [dx].\end{align}
Note that, using \eqref{normcond}, we have 
\begin{align}
	\langle [x],t_0|f(t_0)\rangle=\int f[x']\langle [x],t_0|[x'],t_0\rangle [dx']=f[x].\label{fnorm}
\end{align}
We can then directly interpret 
$f[x]$ as a probability density of Nambu particles 
in the ensemble at an initial time $t=t_0$, 
provided that the 
normalization condition
\begin{align}
\label{stanormal}
	\int f[x]
	[dx]=\iint f[x']\langle [x],t_0|[x'],t_0\rangle [dx'][dx]=1
\end{align}
is satisfied: 
$f[x]$ gives the probability 
distribution of $N$ Nambu particles with respect to the
`complexion' $[x]=(x_1^i,x_2^i,\ldots,x_N^i)\,\, (i=1,2,3)$  of their phase-space coordinates at time $t=t_0$.

It should be clear by making comparison of \eqref{fnorm} with the normalization 
condition \eqref{normcond} that, for a generic statistical state 
$|f(t_0)\rangle$, the precise and maximum information embodied by 
the basis states $|[x],t_0)\rangle$ is 
diminished by the statistical superposition which necessarily 
randomizes microscopic information.  
Furthermore, it should also be clear by construction (and its probability interpretation hitherto given) that 
there is {\it no} interference effect, because the superpositions 
are {\it restricted by definition} with the condition that {\it only 
non-negative} coefficient functions $f[x]$ are allowed. This is a crucial 
feature of our formalism of a classical statistical field 
theory of Nambu dynamics, which the reader is 
required to keep in mind throughout the present paper; even though 
we use the concept of the Liouville field operators 
as formal tools, our interpretation with respect 
to probability associated with state vectors makes a stark contrast to the situation in quantum mechanics, where the normalization 
condition is set for the absolute square of the 
coefficient function. 
Another related comment relevant here is that 
the above classical probability interpretation excludes  
the possibility of a `fermionic' Liouville  field that satisfy  
anti-commutation relation instead of the commutation relations 
\eqref{fcommu}, since clearly a non-negative coefficient function 
$f[x]$ cannot be anti-symmetric under 
exchanges of the coordinates.

\subsection{Further properties of the statistical states}
Even if the distribution function $f[x]$ itself is independent of time $t$, 
the statistical state \eqref{genesta} 
in general has a nontrivial time dependence, 
because of the microscopic streaming of Nambu particles which 
always obey the Nambu equations of motion: using \eqref{nbodycomp} and \eqref{liouvilltheorem}, 
we can write
\begin{align}
	\partial_{t_0}|f(t_0)\rangle &=\hat{{\cal H}}_0|f(t_0)\rangle \nonumber
	\\
	&=
	\int \sum_{a=1}^N
	\bigg(X^i(x_a)\frac{\partial f[x]}{\partial x_a^i}
	\bigg)|[x],t_0\rangle 
	[dx]
	\label{tderfN}.
\end{align}
It is important to notice that 
\begin{align}
	&\int\langle [x],t_0|
	 \partial_
	 {t_0}|f(t_0)\rangle [dx]=\int\sum_{a=1}^N
	\bigg(X^i(x_a)\frac{\partial f[x]}{\partial x_a^i}
	\bigg) [dx]\nonumber \\
	&=-\int\sum_{a=1}^N
	\bigg(\partial_iX^i(x_a)f[x]
	\bigg) [dx]=0.
	\nonumber
\end{align} 
More generally, the Liouville equation 
ensures that the integrated $N$-body operators themselves with the flat  
distribution function and hence the integrated state
$
	\displaystyle \int	\langle [x],t|[dx]
$
and its conjugate,  
are independent of time $t$:
\begin{align}
\frac{d}{dt}\int	\langle [x],t|[dx]=0=
\frac{d}{dt}\int	|[x],t\rangle [dx].	 \label{Fcond}
\end{align}
Although the integrated states themselves are not normalizable 
since they 
correspond to the constant distribution function $f[x]=1$, 
\eqref{Fcond} is meaningful in a formal sense, since 
scalar products with any normalizable statistical states 
$|f(t_0)\rangle$ or $\langle f(t_0)|$ are supposed to be well-defined.

Thus we can represent the normalization condition 
\eqref{stanormal} equivalently by
\begin{align}
	\int	\langle [x],t||f(t_0)\rangle[dx]=1\nonumber
\end{align}
for {\it arbitrary} $t$ and $t_0$. This expresses the 
conservation law for classical probability distribution introduced 
above, since the matrix element $\langle [x],t|f(t_0)\rangle$ is 
interpreted as a 
distribution at an arbitrary time $t$, given the
 initial condition represented by $f[x]$ at time $t_0$. 
For convenience, we designate the integrated $N$-particle state 
with the flat distribution by a special symbol as 
\begin{align}
	\langle Z|\equiv \int \langle [x],t|[dx]\equiv \int \langle [x][dx],\qquad  |Z\rangle \equiv \int | [x],t\rangle [dx] \equiv \int \langle [x]|[dx]	 \nonumber
\end{align}
which are to be called the `Z-vacuum', satisfying by definition  
\begin{align}
	\langle Z|f(t_0)\rangle=1, \nonumber
\end{align}
for an arbitrary statistical state $|f(t_0)\rangle$. 
With an arbitrary physical operator
	$\hat{G}(t)$
which is expressed as a functional of the 
field operators at any time $t$, its expectation value for a 
statistical state $|f(t_0)\rangle$ is now 
given by
\begin{align}
	\langle \hat{G} \rangle_{f(t_0)} (t)=\langle Z|\hat{G}(t)|f(t_0)\rangle.\nonumber
\end{align} 

In fact, we can express the evolution of a generic statistical state at a later time $t>t_0$ explicitly as
\begin{align}
\label{matele}
	\langle [y],t|f(t_0)\rangle =
	\int \sum_{P(a)} f[x]
	\frac{1}{N!}\bigg[\prod_{a=1}^N
	\delta^3\big(y_{P(a)}-x(t-t_0;x_a)\big)\bigg][dx]. 
\end{align}
Owing to the volume-preserving property of time 
evolution, we have the identity
\begin{align}
	\delta^3\big(y-x(t-t_0;x_0)\big)
	=\delta^3\big(x_0-x^{-1}(t-t_0;y))\big)\nonumber
\end{align}
where $x^{-1}(t-t_0;y)$ represents the inversion 
of the equation $y=x(t-t_0;x_0)$, being 
obtained uniquely by solving $x_0$ in terms of $y$. 
Therefore \eqref{matele} is written 
equivalently as 
\begin{align}
	\label{matele2}
	&\langle[y],t|f(t_0)\rangle =\frac{1}{N!}\int \sum_{P(a)} f[x]
	\bigg[\prod_{a=1}^N
	\delta^3\big(x_a-x^{-1}(t-t_0;y_{P(a)})\big)\bigg][dx]	\nonumber\\
	&=f\big(x^{-1}(t-t_0;y_1),x^{-1}(t-t_0;y_2),
	\ldots, x^{-1}(t-t_0;y_N)\big).
\end{align}
In particular, when the initial distribution takes a 
factorized form as 
$f[x]=\prod_{a=1}^N f(x_a)$
in terms of a single function $f(x)$ satisfying $\int f(x)dx=1$, 
we have 
\begin{align}
	\langle [y],t|f(t_0)\rangle
	=\prod_{a=1}^Nf\big(x^{-1}(t-t_0;y_a)\big).\nonumber
\end{align}
These matrix elements are generalizations of the single-body 
Green function $G_{\mathrm{r}}(x,t;x_0,t_0)$ to $N$-body case 
with a statistical 
average over the initial conditions with a given 
distribution function $f[x]$ contained in 
$|f(t_0)\rangle$, and as such 
satisfy also the `many-body' equations of motion,
\begin{align}
	\bigg(\frac{\partial}{\partial t}+\sum_{a=1}^NX^i(y_a)\frac{\partial}{\partial y^i_a}\bigg)\langle[y],t|f(t_0)\rangle=0,\nonumber
\end{align}
with the initial condition
\begin{align}
	\lim_{t\rightarrow t_0}\langle [y],t|f(t_0)\rangle
	=f[y].\nonumber
\end{align}

When \eqref{tderfN} vanishes, namely, the following 
`constraint' is satisfied, 
\begin{align}
\label{stationary}
	\hat{{\cal H}}_0|f(t_0)\rangle=0, 
\end{align}
we say that the ensemble corresponding to the 
statistical state $|f(t_0)\rangle$ is 
stationary. Namely, in that case, $|f(t_0)\rangle$ is 
actually independent of $t_0$. 
Obviously, a set of the 
distribution function $f[x]$ of the following
form 
\begin{align}
	f^{(0)}[x]\equiv g\big(H(x_1),H(x_2),\ldots, H(x_n);
K(x_1),K(x_2),\ldots,K(x_n)\big)\nonumber
\end{align}
which has dependence on the coordinates $x_a$ of the Nambu particles 
composing the ensemble 
only through the energy functions $H$ and $K$, gives 
a stationary ensemble, 
since 
$
X^i(x_a)\partial_{x^i_a}f^{(0)}[x]=0
$
for each $a$ with an arbitrary function $g$ of 
$2N$ variables.

Now, it should be clear that, as long as we remain within the realm of  the usual Nambu equations of motion 
governed by the Hamiltonian $\hat{{\cal H}}_0$, there is no possibility of {\it dynamical} mechanism for attaining 
equilibrium distributions, as we have already stressed in the Introduction. 
Thus, an important next issue is that to what extent it is possible 
set up non-trivial kind interactions for Nambu particles 
consistently, 
such that an ensemble 
may reach an equilibrium statistical state, starting with a class of 
initial states which can be, most typically,   
a stationary statistical state with the distribution function $f^{(0)}[x]$ of the 
above general form. We are now ready to proceed into this problem 
on the basis of general apparatus hitherto constructed.

\section{A Model of Non-Local Interaction}

\subsection{Dynamical evolution of classical statistical states as a Markov process}
A time-dependent 
statistical $N \,(\ge 2)$-body state after including interaction will be 
denoted 
by $|F(t)\rangle$ from now on. The initial statistical state is assumed to be 
\begin{align}
	|F(0)\rangle=
	\int f^{(0)}[x]|
[x],t_0\rangle [dx],\nonumber
\end{align}
which is actually stationary with respect to the free Hamiltonian $\hat{\mathcal H}_0$, i.e. independent of $t_0$, satisfying
\[\hat{\mathcal H}_0|F(0)\rangle=0. \]
For definiteness and notational brevity, 
we set $t_0=0$ and denote the basis 
$N$-body state by
$|[x],0\rangle=|[x]\rangle$. 
A basic {\it premise} for investigating the dynamics of $|F(t)\rangle$ 
 with interaction is that the time evolution 
is described 
as 
\begin{align}
\label{infinidev}
	|F(t+\Delta t)\rangle=e^{\Delta t\hat{{\mathcal H}}_I(t)}|F(t)\rangle \simeq (1+\Delta t \hat{{\mathcal H}}_I(t))
	|F(t)\rangle 
\end{align}
for infinitesimally small $\Delta t$, 
with some interaction Hamiltonian $\hat{{\mathcal H}}_I$ 
which is to be defined {\it independently} of statistical states $|F(t)\rangle$ as a realization 
of the intuitive picture of non-local interaction 
discussed in the Introduction. Thus, 
we have a simple linear differential equation of first order with respect to time:
\begin{align}
\label{diffeqK}
	\partial_t|F(t)\rangle=\hat{{\mathcal H}}_I(t)|F(t)\rangle.
\end{align}
In other words, we assume that the dynamics 
of non-local interaction which is supposed to cause 
the change of the distributions of stationary statistical states is treated as a `Markov process' in continuum time in the space of classical statistical states, if we use the terminology 
of statistical physics.  

Therefore, we now have to treat a time-dependent 
distribution function, $$F([x],t)\equiv F(x_1,\ldots, x_n,t),$$ instead of 
the time-independent distribution function $f[x]$ contained in $|f(t_0)\rangle$ of the 
previous section: namely, we can set
\begin{align}
	|F(t)\rangle=
	\int F([x],t)
	|[x]\rangle [dx],\nonumber
\end{align}
with the initial condition $F([x],0)=f^{(0)}[x]$. 
For convenience, we will call \eqref{diffeqK} the `master equation'. 
In our statistical field theory, 
the master equation plays 
the role of Schr\"{o}dinger 
equation in quantum mechanics. 

Since the interaction of our interest involves a 
certain non-locality 
which 
cannot be dealt with within the framework of the Liouville equation,  
we may call the above operator $\hat{\mathcal H}_I$ 
`stochastic' interaction Hamiltonian. 
However, to avoid possible confusion, we emphasize that, in our case, the `stochasticity' is fundamental and 
intrinsic in a {\it closed} system of $N$ Nambu particles; in other 
words, the Markov process described by \eqref{diffeqK} is autonomous in the sense that  
it is 
{\it not} caused by any external agent, such as a 
heat bath as in the conventional stochasticity   
which are familiar in the usual stochastic dynamics of open systems. 
As a matter of course, if we focus our attention only 
to a single Nambu particle in any closed many-body Nambu systems, 
the former can be regarded effectively as an open system immersed in the latter. 
A simple 
example of such effective approaches  
will be treated briefly for the purpose of comparison with the present formalism 
in Appendix A. 
Remember that, after all, the concept of heat bath 
itself must be based on the existence of interactions at a more 
fundamental level. 

For simplicity, we consider a two-body self-interaction of the following type,
\begin{align}
	\hat{\mathcal{H}}_I(t)=\frac{1}{4}\iiiint\psi^{\dagger}(x_3,t)\psi^{\dagger}(x_4,t)
	V(x_3,x_4;x_1,x_2)\psi(x_1,t)\psi(x_2,t)\prod_{a=1}^4d^3x_a
\label{intH}
\end{align}
where the kernel function $V(x_3,x_4;x_1,x_2)$ is independent of time and is assumed to be real 
and symmetric 
under interchanges $x_1\leftrightarrow x_2, x_3
\leftrightarrow x_4$ and $(x_1,x_2)
\leftrightarrow (x_3,x_3)$ of the coordinates, the last of which 
means that transitions caused by stochastic 
interaction are reversible (a property, usually called {\it microscopic reversibility} in the theory of Markov processes). 
Hence the interaction Hamiltonian is 
hermitian $\hat{\mathcal H}_I
=\hat{\mathcal H}_I^{\dagger}$, in contrast to anti-hermiticity of 
the free Hamiltonian $\hat{{\mathcal H}}_0=-\hat{{\mathcal H}}_0
^{\dagger}$, yielding (real) symmetric matrix elements with respect to the 
basis vectors:
\begin{align}
	\langle [x]|\hat{\mathcal{H}}_I(t)|[y]\rangle=\langle [y]|\hat{\mathcal{H}}_I(t)|[x]\rangle.\label{realsym}
\end{align} 

It should be kept in mind 
that, by definition, the time dependence of the Liouville 
field operators, $\psi(x,t), \psi^{\dagger}(x,t)$, is always governed by 
the original free-field equations, 
the Liouville equations, \eqref{ffeq} in 
our formalism. 
In that sense, even though we are treating a
genuinely a classical statistical system with the 
corresponding classical probabilistic interpretation as formulated 
in section 4, 
our formalism is 
close to the 
so-called `interaction 
representation' which is familiar in perturbative 
{\it quantum} field theories. This essentially reflects our 
intuitive picture of stochastic interactions that, except for 
instantaneous self-interactions shuffling 
initial conditions randomly, each 
of Nambu particles obeys  
the Nambu equations of motion with their own 
unique trajectories, {\it piecewisely} in the 
Nambu phase-spacetime to any finite orders  
of interactions.

It is to be noticed also that we could formally start out with 
the `Schr\"{o}dinger representation' in writing down 
the master equation, instead of the 
interaction representation:
\begin{align}\label{schrep}
	\partial_t|\tilde{F}(t)\rangle =-\hat{{\mathcal H}}
	|\tilde{F}(t)\rangle, 
	\quad 
	\hat{{\mathcal H}}\equiv \hat{{\mathcal H}}_0-\hat{\mathcal{H}}_I,
\end{align}
where the total Hamitonian $\hat{{\mathcal H}}$
is {\it non}-hermitian, $(\hat{{\mathcal H}}_0-\hat{\mathcal{H}}_I)^{\dagger}=-\hat{{\mathcal H}}_0-\hat{\mathcal{H}}_I $, 
and defined in terms of the {\it time-independent} Liouville field operators $\psi(x,0), 
\psi^{\dagger}(x,0)$. 
Then by making a similarity (actually also unitary) transformation
\begin{align}
	|\tilde{F}(t)\rangle =e^{-\hat{{\mathcal H}}_0t}|F(t)\rangle,\label{unitran}
\end{align}
\eqref{schrep} is rewritten 
\begin{align}
	e^{\hat{\mathcal H}_0t}\partial_t \big(e^{-\hat{{\mathcal H}}_0t}|F(t)\rangle
	\big)=-e^{\hat{\mathcal H}_0t}(\hat{{\mathcal H}}_0-\hat{\mathcal{H}}_I)e^{-\hat{{\mathcal H}}_0t}|F(t)\rangle
	\nonumber
\end{align}
which reduces to the master equation \eqref{diffeqK} with, 
\begin{align}
	\hat{{\mathcal H}}_I(t)=e^{\hat{\mathcal H}_0t}
	\hat{\mathcal{H}}_Ie^{-\hat{\mathcal H}_0t}.\nonumber
\end{align}
This is consistent with our original definition of 
basic time-dependent field operators,
$\psi(x,t)=e^{\hat{\mathcal H}_0t}
	\psi(x,0)e^{-\hat{\mathcal H}_0t}, 
\psi^{\dagger}(x,t)=e^{\hat{\mathcal H}_0t}
	\psi^{\dagger}(x,0)e^{-\hat{\mathcal H}_0t}	$. 
Note also that the transformation \eqref{unitran} does not 
violate the non-negativity of the distribution function, 
since the action of the operator $e^{-\hat{\mathcal H}_0t}$ simply induces 
the Nambu equations of motion for the 
probability distribution embodied in $|F(t)\rangle$, 
as is clear from the discussions of section 3.

Finally, we stress that, despite microscopic reversibility, \eqref{realsym},
of the hermitian interaction operator $\hat{\mathcal H}_I(t)$, 
the master equation \eqref{diffeqK} itself {\it cannot} be  
time-reversal invariant in general, just as in the case of 
standard diffusion equation. 
This is in contrast to the Schr\"{o}dinger equation: in the latter case, 
the invariance under time reversal $t\rightarrow -t$ is achieved by complex conjugation 
of the complex wave function, due 
to the presence of imaginary unit on its left hand side 
which is of course absent in \eqref{diffeqK} that 
deals with the real non-negative distribution functions directly. 

\subsection{Requirements for the stochastic interaction}
In the present paper, we restrict ourselves to studying 
a simplest but non-trivial Markov process which can describe 
the evolution of statistical states to equilibrium 
states. 
We require further conditions that 

\vspace{0.1cm}
\noindent
{\bf i}) {\bf Homogeneity}: 
the Markov process of our interest is homogeneous with respect 
to time:
\begin{align}
	\frac{d}{dt}{\hat{\mathcal{H}}_I}=
	[\hat{\mathcal H}_0,\hat{\mathcal H}_I]=0.\nonumber
\end{align}
Because of the field equations \eqref{ffeq}, this is ensured by 
assuming that the kernel function 
$V$ depends on the phase-space coordinates  
essentially only through the energy functions, 
$H(x_a)$ and $K(x_a)$. Consequently,  
the time parameter of the field operators inside of \eqref{intH} 
can be set to
an arbitrary value, say, zero;



\vspace{0.1cm}
\noindent
{\bf ii}) {\bf Non-negativity}: the off-diagonal 
matrix elements of $\hat{\mathcal{H}}_I$ must be {\it non-negative}, 
since the distribution functions cannot be negative at any times. 
That this must be so is easily seen by 
considering infinitesimal time development \eqref{infinidev}.
In contrast to the off-diagonal matrix elements, the signs of diagonal matrix elements of $ 1+\Delta t \hat{{\mathcal H}}_I(t)$ are dominated by 
the first term (i.e., identity operator) on the parenthesis on its right-hand side, and consequently any requirement with respect 
to sign of the diagonal matrix elements is not needed;

\vspace{0.1cm}

\noindent
{\bf iii}) {\bf Conservation of probability}:  to be consistent with the conservation of probability,
\begin{align}
\label{consproZ}
	\langle Z|\hat{{\mathcal H}}_I=0=\hat{{\mathcal H}}_I|Z\rangle,\end{align}
since we must have
$
\langle Z|e^{t\hat{\mathcal{H}_I}}|F(0)\rangle=1
$ for arbitrary initial state $|F(0)\rangle$ at any time $t$. Equivalently, 
\begin{align}
\langle Z|\hat{{\mathcal H}}_I|F(t)\rangle=
\int 
\langle [x]|	 \hat{\mathcal{H}}_I|F(t)\rangle [dx]=0,\nonumber
\end{align}
where for definiteness the time variable 
of the $Z$-vacuum is set at $t=0$, 
remembering that $Z$-vacuum represented as a Fock state constructed 
through the operation of field operators 
are ensured to be independent of time;

\vspace{0.1cm}
\noindent
{\bf iv}) {\bf Conservation of energies}: the conservation of two independent kinds of energies 
separately,
\begin{align}
\label{commuHKint}
	[\hat{{\mathcal H}}_I, \hat{H}]=0
	=[\hat{{\mathcal H}}_I,\hat{K}],
\end{align}
where $\hat{H}$ and $\hat{K}$ are defined by \eqref{HKhat}. 

\vspace{0.2cm}
Now, in order to fulfill {\bf ii}), it is sufficient to 
require that 
\begin{align}
\label{nonnegative}
	V(x_1,x_2;x_3,x_4)\ge 0 \qquad \mbox{for} \qquad (x_1,x_2)\ne 
(x_3,x_4).
\end{align} 
For {\bf iii}), we must have 
\begin{align}
\label{iiiV}
	\iint V(x_3,x_4;x_1,x_2)d^3x_3d^3x_4=0=\iint V(x_3,x_4;x_1,x_2)d^3x_1d^3x_2.
\end{align}
Note that \eqref{nonnegative} and \eqref{iiiV} 
are {\it not} incompatible: they only suggest some 
$\delta$-function-like behavior for 
the diagonal matrix elements. 
For {\bf iv}), we need 
\begin{align}
\begin{aligned}
\label{consV}
	V(x_3,x_4;x_1,x_2)\big(H(x_1)+H(x_2)-H(x_3)-H(x_4)\big)=0,\\
	V(x_3,x_4;x_1,x_2)\big(K(x_1)+K(x_2)-K(x_3)-K(x_4)\big)=0.
	\end{aligned}
\end{align}

These conditions are invariant under 
the linearized form of the $\mathcal{N}$-symmetry transformations, 
namely, SL(2,$\mathbb{R}$) transformations \eqref{sl2r} which is 
globally defined independently of the phase-space coordinates. 
Hence it is guaranteed that the stochastic interaction can be 
formulated in conformity with both the $\mathcal{N}$-symmetry and the gauge symmetry under \eqref{fgaugetrans} with linear $\lambda$ functions, simultaneously. 
It is fairly obvious that, as long as we require the conservation laws for linear 
sums of the energy functions, it is impossible 
to extend the $\mathcal{N}$-symmetry to a fully nonlinear 
form, once we include interactions. In other words, 
the SL(2,$\mathbb{R}$) is essentially the {\it maximal} possible 
symmetry group with respect to the transformations of the 
set of energy functions, which may be imposed upon nontrivial interacting Nambu dynamics.

\subsection{The H-Theorem for the evolution of statistical states}

Next, we demonstrate that the time-evolution governed by 
the master equation \eqref{diffeqK}, with 
the general properties hitherto given, satisfies a version 
of the H-theorem, reflecting the 
irreversibility of the master equation. This is important to us, because it shows a characteristic feature 
of equilibrium statistical states in our formalism. 
To author's knowledge, for standard Markov processes with finite number ($r$) of 
states with discrete time sequences, the 
H-theorem was originally established by Husimi\cite{husimi} (and later 
also by Stueckelberg\cite{stueck}). 
Fortunately, his argument 
can be extended straightforwardly to our case. 

As a preparation, let us briefly recapitulate Husimi's proof. 
The basic condition required is that the transition 
probability $p_{kj}(\Delta t)\, (\ge 0)$ corresponding to 
the transition from a state $j$ to a state 
$k $ for a small time-interval $\Delta t$ satisfies the normalization 
conditions with respect to {\it both} indices $k,j$:
\begin{align}
	\sum_{k=1}^rp_{kj}(\Delta t)=1 \qquad \text{and}\qquad \sum_{j=1}^rp_{kj}
	(\Delta t)=1.\label{normalcond2}
\end{align}
The first equality comes from the 
definition of transition probability $p_{kj}(\Delta t)$ itself as usual. 
On the other hand, the second one is satisfied automatically 
as a consequence of the first {\it if} we assume microscopic reversibility, 
namely 
$p_{kj}(\Delta t)=p_{jk}(\Delta t)$. 
It is important 
to notice that the latter equality of \eqref{normalcond2} guarantees 
that $p_{kj}(\Delta t)$ provides a role of 
some probability distribution, denoted by Pr-I,
\begin{align}
	\big (p_{k1}(\Delta t),p_{k2}(\Delta t),
	\ldots, p_{kr}(\Delta t)
\big)
\label{tdist}
\end{align}
with respect to the running index $j$ with fixed $k$, 
in addition to the distribution, denoted by Pr-0,
$$
\big(p_{1j}(\Delta t),p_{2j}(\Delta t),\ldots,p_{rj}(\Delta t)\big)
$$ with respect to the running index $k$ with fixed $j$ as 
in the case of first equality of \eqref{normalcond2}. 
Consider the stochastic distribution 
functions $\big(f_1(t),f_2(t),\ldots, f_r(t)\big)$ at time $t$, 
satisfying by definition, 
\[
f_i(t)\ge 0, \qquad \sum_{i=1}^rf_i(t)=1,
\] 
whose evolution 
is governed by the transition probability $p_{kj}(\Delta t)$.   
Thus, at $t+\Delta t$, distribution functions  
are given by $\displaystyle f_k(t+\Delta t)=\sum_{j=1}^rp_{kj}(\Delta t)f_j(t)$. 
We denote the expectation value of any function $h(f)$ 
with respect to the probability distribution Pr-I, \eqref{tdist},
\begin{align}
	\langle h(f) \rangle_k\equiv \sum_jp_{kj}h(f_j).\nonumber
\end{align}
If $h(f)$ is chosen to be a (downward) convex function, we have 
the well-known inequality
\begin{align}
	h(\langle f \rangle_k)\le \langle h(f) \rangle_k, \nonumber
\end{align}
which is easily proven graphically. 
Thus, we have, in the sense of the distribution Pr-I
\begin{align}
	h\big(\sum_jp_{kj}(\Delta t)f_j(t)
	\big) \le \sum_jp_{kj}(\Delta t)h(f_j(t)).\nonumber
\end{align}
By taking the sum over the remaining index $k$ on 
both sides, we obtain, using the definition of Pr-0, 
\begin{align}
	\sum_kh\big(f_k(t+\Delta t)\big)=\sum_kh\big(\sum_jp_{kj}(\Delta t)f_j(t)
	\big) \le \sum_j h\big(f_j(t)\big).\nonumber
\end{align}
This shows that the `H-function' 
\begin{align}
	{\mathsf H}(t)\equiv \sum_kh(f_k(t))\nonumber
\end{align}
can only decrease, or remain constant. The latter case 
occurs for sufficiently large $t$ when the evolution 
reaches an equilibrium, provided that 
$\big(f_1(\infty),f_2(\infty),\ldots, f_r(\infty)\big)$ is well-defined. In particular, if we 
choose $h(f)=f\log f$, $-{\mathsf H}$ in equilibrium is essentially 
the definition of entropy $S$, apart from the
Boltzmann constant: 
\begin{align}
	-{\mathsf H}=S=-\sum_k f_k\log f_k.
	\label{Hfuncstand}
\end{align}
It is important to notice that the 
assumptions which are essential in this argument are only the conditions  \eqref{normalcond2} which are 
independent of the details of the dynamical mechanism of 
transitions: this is 
in sharp contrast to the well-known Boltzmann's H-theorem\footnote{For a comprehensive modern account of the H-theorem in the framework of the standard statistical mechanics, see, e.g., the reference.\cite{haar}} in the case 
of gas theory. In the 
latter, 
further assumptions must be invoked, such as the 
`Stosszahlenansatz' (often called `scattering assumption' of 
`molecular chaos') 
for collision of gas molecules.

Now, in our case of the Markov process governed by the 
master equation, 
all necessary assumptions 
are met, except for a difference that we are treating 
a continuously infinite number of states with continuous 
indices denoted by bra-vector $\langle [x]|$ and 
continuous state denoted by ket-vector $|F(t)\rangle$. 
We adopt the H-function of the above form 
\eqref{Hfuncstand} extended to continuous case:
\begin{align}
	h(F(t))
	=\langle [x]|F(t)\rangle \log \langle [x]|F
	(t)\rangle.
	\nonumber
\end{align}
A rationale 
for this particular choice is that it 
naturally satisfy the following requirement: when we consider 
two systems ${\mathcal S}_1$ and ${\mathcal S}_2$ which are completely independent to each other with separate basis states $|[x_1]\rangle$ and $|[x_2]\rangle$,  
the statistical state $|F_{12}(t)\rangle $ of the combined system ${\mathcal S}=
{\mathcal S}_1 +\mathcal{S}_2$ takes 
the form of a direct product 
$$\langle [x_1]|\langle [x_2]|F_{12}(t)\rangle \equiv \langle [x_1]|F_1(t)\rangle  \langle [x_2]|F_2(t)\rangle, 
\quad 
\int \langle [x_a]|F(t)\rangle [dx_a]=1, \quad (a=1,2).
$$ Then the  
H-function of the combined system should be a direct sum of the H-functions of the two systems for arbitrary $t$:
\begin{align}
	{\mathsf H}_{12}(t)=\int \langle &[x_1]|\langle [x_2]|F_{12}(t)\rangle \log \big\{\langle [x_1]|\langle [x_2]|F_{12}
	(t)\rangle \big\}\, [dx_1][dx_2]={\mathsf H}_1(t)+{\mathsf H}_2(t), \nonumber \\
	&{\mathsf H}_a(t)=\int \langle [x_a]|F_a(t)\rangle \log \langle [x_a]|F_a
	(t)\rangle [dx_a], 
	\quad (a=1,2). \nonumber
\end{align}
The significance of this trivially looking property lies in that it 
corresponds to the 
`extensiveness' of entropy: in our 
case of Nambu dynamics, we do not have the clear 
concept of `volume' and hence neither of a `box' of a finite  
volume, enclosing our system 
of $N$ Nambu particles by which we usually 
express the extensiveness, since we do not, in principle, have 
any clear distinction between canonical coordinates and 
the conjugate momenta. Remember also that the 
concept of volume itself presupposes the interaction 
of the system with environment (as the `wall') that is basically 
absent in the 
case of Nambu dynamics, at least at the outset,
 as we have stressed in the 
Introduction.

The transition probability, corresponding to the above $p_{kj}$,  
from a state function $\langle [x]|F(t)\rangle$ 
(in place of $f_j(t)$) to a state function 
$\langle [x]|F(t+\Delta t)\rangle$ (in place of $f_k(t+\Delta t)$) is given 
by the matrix element
\begin{align}
	p([x],[y])\equiv \langle [x]|e^{\Delta t \hat{\mathcal{H}}_I}|[y]\rangle
	=\langle [y]|e^{\Delta t \hat{\mathcal{H}}_I}|[x]\rangle
	\equiv p([y],[x]),\nonumber
\end{align}
which is ensured to be symmetric and non-negative 
due to our requirement {\bf ii}). 
We have, using the completeness relation \eqref{comprelation}, 
\begin{align}
	\langle [x]|F(t+\Delta t)\rangle=\int \langle [x]|
	e^{\Delta t \hat{\mathcal{H}}_I}|[y]\rangle\langle [y]|F(t)\rangle [dy].\nonumber
\end{align}
Note also that \eqref{consproZ}, the requirement {\bf iii}), guarantees that
\begin{align}
	\int p([x],[y]) [dx]=1=\int p([x],[y])[dy].\nonumber
\end{align}
Hence we safely obtain 
the H-theorem for the time evolution of $|F(t)\rangle$
governed by our master equation, 
\begin{align}
	{\mathsf H}(t+\Delta t)=\int h\big(\langle [x]|F(t+\Delta t)\rangle\big)[dx]
	\le \int h\big(\langle [x]|F(t)\rangle\big)[dx]=
	{\mathsf H}(t).\nonumber
\end{align}
Thus an equilibrium state $|F(\infty)\rangle$, 
satisfying
$
	\hat{\mathcal{H}}_I|F(\infty)\rangle=0\nonumber
$,
must minimize
the H-function:
\begin{align}
	{\mathsf H}(\infty)=\int h\big(\langle [x]|F(\infty)\rangle\big)[dx]
	=\int \langle [x]|F(\infty)\rangle \log \langle [x]|F
	(\infty)\rangle [dx]
	\label{entropyN}
\end{align}
among all possible states that are connected through the master 
equation: here, it is important to keep in mind that the nature of equilibrium 
states $|F(\infty)\rangle$ in general depend on the initial state
$|F(0)\rangle$ and hence on conditions chosen for  
the initial distribution function $f^{(0)}[x]$. 

This minimum principle (essentially, `the principle of entropy' as in the usual statistical mechanics), which is derived as a 
consequence of dynamics governed by the master equation 
from the H-theorem, will 
play an indispensable role for characterizing the equilibrium states 
of statistical Nambu dynamics in the next section: it 
takes the place of the variational 
principles (which characterize the Nambu equations of motion 
on the basis of Stokes' theorem) in {\it non}-statistical Nambu dynamics as elucidated in\cite{yoneintro}.

\subsection{Ansatz for stochastic interaction}
We now propose a simple model 
for the stochastic interaction which is viable for 
a reasonably concrete discussion on the approach to 
equilibrium statistical states. 
First we rewrite the condition \eqref{iiiV} by 
redefining the kernel function as 
\begin{align}
\label{totalkernel}
	V(x_3,x_4;x_1,x_2)\equiv g^2\big(v(x_3,x_4;x_1,x_2)
	-i(x_3,x_4;x_1,x_2)\big)
\end{align}
where $g^2$ is a positive coupling constant of engineering dimension 
[time]$^{-1}$, and 
\begin{align}
	i(x_3,x_4;x_1,x_2)\equiv 
	\frac{1}{2}
	\big(\delta^3(x_3-x_1)\delta^3(x_4-x_2)
	+\delta^3(x_3-x_2)\delta^3(x_4-x_1)\big),\nonumber
\end{align}
satisfying
\begin{align}
	\iint i(x_3,x_4;x_1,x_2)d^3x_3d^3x_4=1=\iint i(x_3,x_4;x_1,x_2)d^3x_1d^3x_2.\nonumber
\end{align}
Then, \eqref{iiiV} is equivalent to 
\begin{align}
\label{iiiv1}
	\iint v(x_3,x_4;x_1,x_2)d^3x_3d^3x_4=1=\iint v(x_3,x_4;x_1,x_2)d^3x_1d^3x_2.
\end{align}
To fulfill the condition \eqref{nonnegative}, we require 
that the reduced kernel function $v(x_3,x_4;x_1,x_2)$ is non-negative. Since the identity function $i(x_3,x_4;x_1,x_2)$ by definition 
satisfies the conservation law \eqref{consV} identically, 
$v$ must also obey
\begin{align}
	\begin{aligned}
\label{consv}
v(x_3,x_4;x_1,x_2)\big(H(x_1)+H(x_2)-H(x_3)-H(x_4)\big)=0,\\
	v(x_3,x_4;x_1,x_2)\big(K(x_1)+K(x_2)-K(x_3)-K(x_4)\big)=0.
\end{aligned}
\end{align}
Thus, the reduced kernel function $v(x_3,x_4;x_1,x_2)$ 
can have nonzero values only when 
$k_{12}=k_{34}, h_{12}=h_{34}$ where 
$k_{ab}\equiv K(x_a)+K(x_b), h_{ab}\equiv H(x_a)+H(x_b)$.

It is not difficult to construct a concrete example for $v$ which 
satisfies all of the above requirements. 
First remember as the most important and crucial characteristic 
of our approach to non-local interaction that the dependence on 
phase-space coordinates must occur only through 
energy functions as we have already stressed in connection with 
the requirement {\bf i}) in subsection {\it 4.1}. For fulfillment of the
conservation laws \eqref{consv}, it is natural to set
\begin{align}
	&v(x_3,x_4;x_1,x_2)=\bar{v}(x_3,x_4;x_1,x_2)
	\nonumber \\
	&\times \delta(H(x_1)+H(x_2)-H(x_3)-H(x_4))\delta(K(x_1)+K(x_2)-K(x_3)-K(x_4)).
\end{align}
In order to obtain a concrete example for
 the coefficient function $\bar{v}$, 
it is convenient to rewrite the product of the delta 
functions formally as
\begin{align}
	&\delta(H(x_1)+H(x_2)-H(x_3)-H(x_4))\delta(K(x_1)+K(x_2)-K(x_3)-K(x_4))\nonumber\\
	&=\iint\iint\delta(H(x_1)+H(x_2)-h_{12})\delta(h_{12}-h_{34})\delta(H(x_3)+H(x_4)-h_{34})\nonumber\\
	&\times \delta(K(x_1)+K(x_2)-k_{12})
	\delta(k_{12}-k_{34})\delta(K(x_3)+K(x_4)-k_{34})dh_{12}
	dh_{34}dk_{12}dk_{34},\nonumber
\end{align}
introducing auxiliary integration variables $(h_{ij},k_{ij})$ 
with $(ij)=(12),(34)$. 
Namely, we decompose the product of $\delta$-functions 
according to the values of the sums of energy 
functions of the initial and final coordinates $(x_1,x_2)$ 
and $(x_3,x_4)$. 
Due to our assumptions on 
the functions $(H(x),K(x))$ 
stated previously, the following integral 
\begin{align}
	\rho_{H,K}(h,k)\equiv \iint \delta(H(x_1)+H(x_2)-h)\delta(K(x_1)+K(x_2)-k)
	d^3x_1d^3x_2
	\label{2bodyrho}
\end{align} 
gives a well-defined function for generic positive 
values of the auxiliary variables $(h,k)$. This is  
non-vanishing only when two level hyper-surfaces 
corresponding to the equality $(H(x_1)+H(x_2),K(x_1)+K(x_2))=(h,k)$ 
have intersections, and consequently its value is proportional to the volume of a compact 
four-dimensional object embedded in 
six-dimensional coordinate space $(x_1,x_2)$
 which increases monotonically for 
large values of $h,k$, with the constraints $H(x_1)+H(x_2)=h, K(x_1)+K(x_2)=k$. It is to be noted that in general the function $\rho_{H,K}(h,k)$ depends on the functional form of the energy functions $H(x),K(x)$: the lower suffix $H,K$ is placed to signify this dependence explicitly.  

Now due to the definition of $\rho_{H,K}(h,k)$, the function $\bar{v}$ can be chosen to be 
\begin{align}
	\bar{v}(x_3,x_4;x_1,x_2)=&[\rho_{H,K}(H(x_3)+H(x_4),
	K(x_3)+K(x_4))\nonumber \\
	&\times \rho_{H,K}(H(x_1)+H(x_2),
	K(x_1)+K(x_2))]^{-1/2}.
\end{align}
Notice that, as this expression appears 
only as the coefficient in front of a product of  
delta-functions
\[
\delta(H(x_1)+H(x_2)-H(x_3)-H(x_4))\delta(K(x_1)+K(x_2)-K(x_3)-K(x_4)),
\]
the function $\bar{v}(x_3,x_4;x_1,x_2)$ can be 
replaced by $\rho_{H,K}\big(H(x_1)+H(x_2),K(x_1)+K(x_2)\big)^{-1}$ 
or $\rho_{H,K}\big(H(x_3)+H(x_4),K(x_3)+K(x_4)\big)^{-1}$ 
depending on various situations. Using this property, 
it is straightforward to check that the conditions \eqref{iiiv1} is satisfied :
\begin{align}
	&\iint v(x_3,x_4;x_1,x_2)d^3x_3d^3x_4\nonumber\\
	&=\iint\rho_{H,K}(H(x_3)+H(x_4);K(x_3)+K(x_4))^{-1}\iint \delta(H(x_1)+H(x_2)-h_{12})\nonumber\\
	&\times \delta(K(x_1)+K(x_2)-k_{12})
	\delta(H(x_3)+H(x_4)-h_{34})
	\delta(K(x_3)+K(x_4)-k_{34})\nonumber \\
	&\times \delta(h_{12}-h_{34})\delta(k_{12}-k_{34}) dh_{12}dk_{12}
	dh_{34}dk_{34}d^3x_3d^3x_4\nonumber \\
	&=\iint\rho_{H,K}(h_{34};k_{34})^{-1}\iint \delta(H(x_1)+H(x_2)-h_{12})\nonumber\\
	&\times \delta(K(x_1)+K(x_2)-k_{12})
	\delta(H(x_3)+H(x_4)-h_{34})
	\delta(K(x_3)+K(x_4)-k_{34})\nonumber \\
	&\times \delta(h_{12}-h_{34})\delta(k_{12}-k_{34}) dh_{12}dk_{12}
	dh_{34}dk_{34}d^3x_3d^3x_4\nonumber \\
	&=\iint \delta(H(x_1)+H(x_2)-h_{12})\delta(K(x_1)+K(x_2)-k_{12})
	\delta(h_{12}-h_{34})\delta(k_{12}-k_{34})
	\nonumber \\
	&\times dh_{12}dk_{12}
	dh_{34}dk_{34}=1, \nonumber
\end{align}
where, in the third equality, we performed 
integrations over the coordinates $(x_3,x_4)$ 
before those over the auxiliary variables $(h_{12},k_{12}, 
h_{34},k_{34})$. 
Since all of the above equations 
are essentially invariant under the SL(2,${\mathbb R}$) 
transformations \eqref{sl2r} provided that 
the auxiliary integration variables $(h_{ij},k_{ij})$ 
transform as the fundamental doublet representation of SL(2,${\mathbb R}$) under which 
the integration measure $dh_{ij}dk_{ij}$ is 
invariant, the result is valid and equivalent for 
any choices of the energy functions $(H,K)$ which are connected by 
the $\mathcal{N}$-symmetry transformations.

\subsection{Case of quadratic 
energy functions}

Since the discussion of the previous subsection is 
somewhat abstract, it will perhaps be meaningful here to 
give an explicit example of the function $\rho(h,k)$ 
in the case of quadratic energy functions:
\begin{align}
	H=A_1(x^1)^2+A_2(x^2)^2+A_3(x^3)^2, \qquad K=B_1(x^1)^2+B_2(x^2)^2+B_3(x^3)^2,\nonumber
\end{align}
where $A_i$ and $B_i$ are two different sets ($A_i\ne B_i$) of positive constants. 
This includes the typical case \eqref{Eulertop} of the rigid rotator. 
We have to compute
\begin{align}
	\rho_{H,K}(h,&k)=
\iint\delta\big(A_1[(x^1_1)^2+(x^1_2)^2]+A_2[(x^2_1)^2+(x^2_2)^2]+A_3[(x^3_1)^2+(x^3_2)^2]-h\big) \nonumber
\\
&\times \delta\big(B_1[(x^1_1)^2+(x^1_2)^2]+B_2[(x^2_1)^2+(x^2_2)^2]+B_3[(x^3_1)^2+(x^3_2)^2]-k\big)d^3x_1d^3x_2.
\nonumber
\end{align}
The integration measure
$d^3x_1d^3x_2=dx^1_1dx^1_2dx^2_1dx^2_2dx^3_1dx^3_2$
can be transformed into that of polar coordinates 
$(x^i_1,x^i_2)=r^i(\cos\theta^i,\sin\theta^i)$ 
$(i=1,2,3)$ for each $i$:
$
dx^i_1dx^i_2=r^1dr^id\theta^i=\frac{1}{2}d(r^i)^2d\theta^i, 
$
and angular integrations give the factor $\pi^3$. 
By making a redefinition $(r^i)^2=R_i$, 
we write
\begin{align}
	\rho_{H,K}(h,k)=\pi^3\underset{R_i\ge 0}{\int}\delta(A_1R_1+A_2R_2+A_3R_3-h)
\delta(B_1R_1+B_2R_2+B_3R_3-k)d^3R.\nonumber
\end{align}
Thus the level surfaces of the energy functions are now 
metamorphosed into 
flat planes, which we call 
`$h$-plane' 
for $A_1R_1+A_2R_2+A_3R_3=h$ and `$k$-plane'  
$B_1R_1+B_2R_2+B_3R_3=k$, both being limited 
in the first octant of the three-dimensional space 
$(R_1,R_2,R_3)$. Note that each point of this 3-dimensional 
space actually represents the 3-dimensional torus $(\theta^1,\theta^2, 
\theta^3)$ $(0\le  \theta^i \le  2\pi)$. 
The intersection 
of the level planes is a straight line, which we call 
`I-line' for convenience, connecting two out of the 
three coordinate planes ($R_i=0, R_j\ge  0$, $j\ne i$, $i,j=1,2,3$) which forms the three sides of the octant. 
The above integral is then proportional to the length of the I-line. 
Of course, the I-line actually represents the 4-dimensional sub-manifold embedded 
in 6-dimensional space corresponding to the pair 
of 3-dimensional coordinates $(x^i_1,x^i_2$), 
corresponding to the conditions $H(x_1)+H(x_2)=h$ 
and $K(x_1)+K(x_2)=k$.  As a typical situation with a nontrivial intersecting line, 
let us consider the case where the following 
conditions for the values of $(h,k)$ are met, either
\begin{align}
	{\rm (I)}: \qquad h/A_1<k/B_1, \quad h/A_2<k/B_2, \quad h/A_3>k
	/B_3, \nonumber
\end{align}
or
\begin{align}
	{\rm (II)}: \qquad h/A_1>k/B_1, \quad h/A_2>k/B_2,\quad h/A_3< k/B_3.
	\nonumber
\end{align}
Other possible cases for the occurrence of nontrivial intersections 
are obtained from these two cases by exchanging the  
indices appropriately. Also the case (II) is obtained 
from (I) by the interchange $h\leftrightarrow k, A_i\leftrightarrow B_i$, so that it is sufficient to treat only the case (I). 
The relevant geometry is illustrated in Fig. 1. 
By performing the integration explicitly, we easily get 
\begin{align}
	\rho_{H,K}(h,k)=\frac{\pi^3|hB_3-kA_3|}{|A_3B_1-A_1B_3||A_3B_2-A_2B_3|}.
	\label{rhoexample}
\end{align}

On the other hand, the coordinates at the ends of the I-line 
on the coordinate planes are 
\begin{align}
	&{\rm 13 \, plane}: \quad (R_1^{(13)},0,R_3^{(13)})\equiv \bigg(\bigg|\frac{-hB_3+kA_3}{A_3B_1-A_1B_3}\bigg|, 0,
\bigg|\frac{hB_1-kA_1}{A_3B_1-A_1B_3}\bigg|\bigg),\nonumber\\
&{\rm 23\, plane}: \quad (0, R_2^{(23)},R_3^{(23)})\equiv \bigg(0, \bigg|\frac{-hB_3+kA_3}{A_3B_2-A_2B_3}\bigg|, 
\bigg|\frac{B_2h-A_2k}{A_3B_2-A_2B_3}\bigg|\bigg).\nonumber
\end{align}
Hence the length of the I-line is equal to 
\begin{align}
	&\big[(R_1^{(13)})^2+(R_2^{(23)})^2+(R_3^{(13)}-R_3^{(23)})^2
\big]^{1/2}\nonumber\\
&=\frac{|hB_3-kA_3|}{|(A_3B_1-A_1B_3)(A_3B_2-A_2B_3)|}\nonumber\\
&\times \big[(A_3B_2-A_2B_3)^2+(A_3B_1-A_1B_3)^2+(A_1B_2-A_2B_1)^2\big]
^{1/2}.\nonumber
\end{align}
Thus apart from a universal factor $\big[(A_3B_2-A_2B_3)^2+(A_3B_1-A_1B_3)^2+(A_1B_2-A_2B_1)^2\big]
^{1/2}$, which is completely symmetric under arbitrary 
interchanges of indices and is independent of the values $(h,k)$ 
of energy functions, $\rho_{H,K}(h,k)/\pi^3$ is essentially the length of the 
I-line in the 3-dimensional space $(R_1,R_2,R_3)$. In the sense of the original 
6-dimensional coordinate space, \eqref{rhoexample} gives the volume of 
the 4-dimensional sub-manifold embedded as 
the intersection of the level hyper-surfaces 
corresponding to $(H(x_1)+H(x_2),K(x_1)+K(x_2))=(h,k)$. 

\begin{figure}
\centering
	\includegraphics[width=7cm]{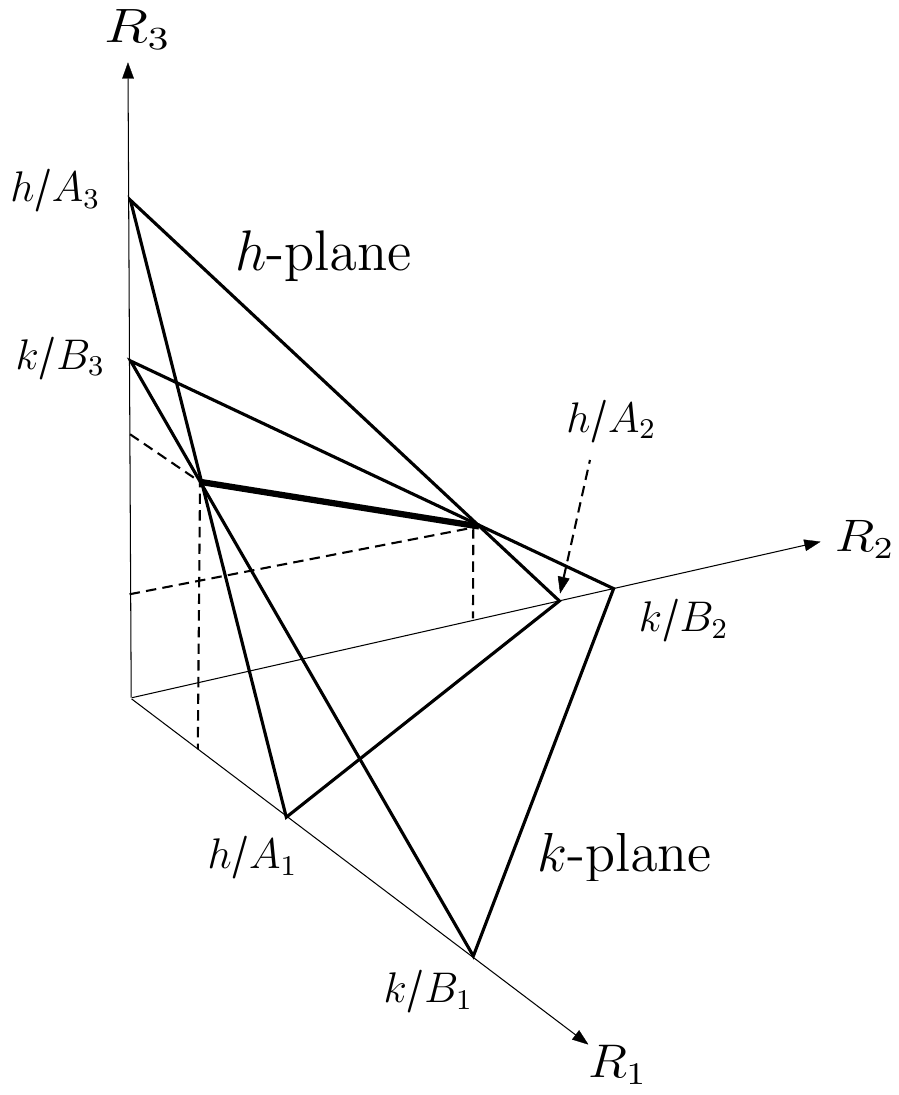}
	\label{Fig.1}
	\caption{The I-line 
	at the intersection of $h$-plane and $k$-plane.}
\end{figure}

We can see clearly how the 
$\mathcal{N}$-symmetry is realized in this example. 
First we note that the 
pairs $(A_i,B_i)$ $(i=1,2,3)$ and $(h,k)$ are 
transformed according to the fundamental doublet 
representation of SL(2,${\mathbb R}$). 
The `cross-product' of any two different doublets is 
invariant under the transformation:
\begin{align}
	A_1'B_2'-A_2'B_1'&=(aA_1+bB_1)(cA_2+dB_2)-(aA_2+bB_2)(cA_1+dB_1)\nonumber\\
	&=(ad-bc)(A_1B_2-A_2B_1)= A_1B_2-A_2B_1. \nonumber
	\end{align} 
Therefore, the result \eqref{rhoexample} is manifestly invariant under the $\mathcal{N}$-symmetry transformation:
\[
\rho_{H',K'}(h',k')=\rho_{H,K}(h,k),
\]
 since all 
of these doublets appear only through the form of 
the cross-products, $A_iB_j-A_jB_i$ $\big(i\ne j; i,j\in 
(1,2,3)\big)$ 
and $hB_i-kA_i$ $\big(i \in (1,2,3)\big)$. Note also that the absolute values of these cross-products are invariant under the simultaneous 
interchanges $A_i\leftrightarrow B_i$ and $h\leftrightarrow k$. 
Hence our conclusion with respect to the 
$\mathcal{N}$ symmetry is valid also in the case II 
as it is.

\subsection{The exact spectrum and eigenfunctions for the functional operator  $v(x_3,x_4;x_1,x_2)$}

We now show that the eigenvalue spectrum of the kernel 
function $v(x_3,x_4;x_1,x_2)$ defined above consists only of two possible values,  
$1$ and $0$, when $v(x_3,x_4;x_1,x_2)$ is regarded as a symmetric and 
real matrix with 
continuous indices, $(x_3,x_4)$ and $(x_1,x_2)$ being the row and column indices, respectively. 
Each eigenvalues are infinitely degenerate due to the presence of continuous values of energy functions. 

The eigenvalue equation is 
an integral equation
\begin{align}
	\iint v(x_3,x_4; x_1,x_2)f_{\lambda}(x_1,x_2)d^3x_1d^3x_2=
	\lambda f_{\lambda}(x_3,x_4),\nonumber
\end{align}
with $f_{\lambda}(x_1,x_2)$ being the eigenfunction. 
Because of the symmetry of the kernel function under the interchange of
the coordinates $x_3\leftrightarrow x_4$ and 
$x_1\leftrightarrow x_2$, we can assume that the 
eigenfunctions are also symmetric, 
$f(x_1,x_2)=f(x_2,x_1)$. Since the kernel function conserves 
the energies, the eigenfunctions can be decomposed linearly 
into 
\begin{align}
	f_{\lambda}(x_1,x_2)=\iint f_{\lambda}(x_1,x_2|k,h)dkdh,
	\nonumber
\end{align}
such that the component function $f_{\lambda}(x_1,x_2|k,h)$ 
can be nonzero only for the coordinates $(x_1,x_2)$ satisfying 
\begin{align}
	k=K(x_1)+K(x_2), \qquad h=H(x_1)+H(x_2)\nonumber
\end{align}
for any given allowed values of $k$ and $h$, and the eigenvalue 
equation is reduced to
\begin{align}
\label{eigenlambda}
	\iint v(x_3,x_4; x_1,x_2)f_{\lambda}(x_1,x_2|k,h)d^3x_1d^3x_2=
	\lambda f_{\lambda}(x_3,x_4|k,h).
\end{align}
Our assumption on the energy functions ensures that 
the integral with respect to 
$(x_1,x_2)$ in this eigenvalue equation is over 
compact manifold of finite volume and hence is well defined. 
By comparing this equation with the condition \eqref{iiiv1}, 
we immediately see that a {\it constant function} 
\begin{align}
	f_1(x_1,x_2|k,h)=f(k,h) \label{constf}
\end{align}
with each given $(k,h)$ 
is an eigenfunction with eigenvalue $\lambda=1$. 

Now suppose there exists an eigenfunction $f_{\lambda}(x_1,x_2|h,k)$ 
with eigenvalue $\lambda$ {\it different} from 1. Then, we can derive by 
integrating the both sides of \eqref{eigenlambda} over $(x_3,x_4)$,
\begin{align}
	\iint f_{\lambda}(x_1,x_2|h,k)d^3x_1d^3x_2=0,\label{firstexcite}
\end{align}
which is nothing but the usual orthogonality condition in disguise. 
On the other hand, for any given $(x_3,x_4)$ 
with any allowable and fixed $k=k_{34}, h=h_{34}$, 
the left-hand side of \eqref{eigenlambda} is by itself proportional 
to $\iint f_{\lambda}(x_1,x_2|h,k)d^3x_1d^3x_2$, since 
$v(x_3,x_4;x_1,x_2)$ is also {\it constant} with respect to $(x_1,x_2)$ under the given condition. 
This necessarily implies $\lambda=0$ owing to the 
orthogonality condition above. This is what we promised to prove. 

The reason why the integral operator $v(x_3,x_4;x_1,x_2)$ 
has such a simple property with respect to the eigenvalue 
spectrum is that it is 
actually a {\it projection} operator. In fact, we can directly check, 
using some of the properties used in the calculations of the 
previous subsection, the following identity:
\begin{align}
	&\iint v(x_3,x_4;y_1,y_2)v(y_1,y_2;x_1,x_2)d^3y_1d^3dy_2\nonumber=\iint \bar{v}(x_3,x_4;y_1,y_2)\bar{v}(y_1,y_2;x_1,x_2)	\nonumber \\
	&\times \delta(H(y_1)+H(y_2)-H(x_3)-H(x_4))\delta(K(y_1)+K(y_2)-K(x_3)-K(x_4))\nonumber \\
	&\times \delta(H(x_1)+H(x_2)-H(y_1)-H(y_2))\delta(K(x_1)+K(x_2)-K(y_1)-K(y_2))	d^3y_1d^3y_2	\nonumber \\
	&=\bar{v}(x_3,x_4;x_1,x_2)^2\delta(H(x_1)+H(x_2)-H(x_3)-H(x_4))\delta(K(x_1)+K(x_2)-K(x_3)-K(x_4))\nonumber \\ 
	&\times \iint 
	\delta(H(y_1)+H(y_2)-H(x_3)-H(x_4))\delta(K(y_1)+K(y_2)-K(x_3)-K(x_4))d^3y_1d^3dy_2\nonumber\\
	&=\bar{v}(x_3,x_4;x_1,x_2)\delta(H(x_1)+H(x_2)-H(x_3)-H(x_4))\delta(K(x_1)+K(x_2)-K(x_3)-K(x_4))	\nonumber\\
	&=v(x_3,x_4;x_1,x_2).\nonumber
\end{align}
The origin of this property is that the conditions \eqref{iiiv1} and 
\eqref{consv} are satisfied by any power of an integral operator 
$v(x_3,x_4;x_1,x_2)$, once they are satisfied by 
a single power of it: this is obvious for 
\eqref{consv}. In the case of \eqref{iiiv1}, we have
\begin{align}
	\iint\hspace{-0.2cm}\iint v(x_3,x_4;y_1,y_2)v(y_1,y_2;x_1,x_2)d^3y_1d^3y_2d^3x_3
	d^3x_4=\iint v(y_1,y_2;x_1,x_2)d^3y_1d^3y_2=1.\nonumber
\end{align} 
Note that the total kernel 
operator $v_{\perp}\equiv v(x_3,x_4;x_1,x_2)-i(x_3,x_4;x_1,x_2)$ is 
also a projection operator, which is orthogonal to 
$v(x_3,x_4;x_1,x_2)$:
\begin{align}
	\iint v_{\perp}&(x_3,x_4;y_1,y_2)v_{\perp}(y_1,y_2;x_1,x_2)d^3y_1d^3dy_2=v_{\perp}(x_3,x_4;x_1,x_2),
	\nonumber\\
	&\iint v(x_3,x_4;y_1,y_2)v_{\perp}(y_1,y_2;x_1,x_2)d^3y_1d^3y_2=0.
	\nonumber
\end{align}
These are remarkable characteristics of our simple model of 
`stochastic' 
non-local interaction. 
As a possible help for the reader to grasp somewhat peculiar 
structure of our kernel function, 
we present a prototypical toy model
in terms of discrete matrices which 
captures, to a certain degree, some critical aspects of the above properties of the kernel function in Appendix B. 

\section{Equilibrium statistical states}

The result of the previous section about our simple of model 
for a stochastic non-local interaction established 
the following fact: the eigenvalue spectrum 
of the total kernel function, $V(x_3,x_4;x_1,x_2)
=g^2(v(x_3,x_4;x_1,x_2)-i(x_3,x_4;x_1,x_2))$,   
 as an integral operator, consists of only two values, $g^2-g^2=0$ and $0-g^2=-g^2$ with an infinite degeneracy in each case. 
In this section, we first discuss 
implications of this result for general many-body statistical states, 
and then proceed to examine the properties of the equilibrium statistical states from a general point of view. 

\subsection{Approaches to equilibrium statistical states}
By acting the interaction operator to the 
basis $N$-body state $|[x]\rangle$, we obtain
\begin{align}
	{\mathcal H}_I(t)|[x]\rangle&=\frac{1}{4\sqrt{N!}}
	\iint
	V(x_3',x_4';x_a,x_b)\psi^{\dagger}(x_3',0)
	\psi^{\dagger}(x_4',0)
	\nonumber\\
	&\times \sum_{a,b\, (a\ne b)}^N
	\bigg(\prod_{c\ne a,b}^N
	\psi^{\dagger}(x_c,0)\bigg)|0\rangle d^3x_3'd^3x_4'.
	\nonumber
\end{align}
This shows that, for the generic $N$-body statistical 
state $|F(t)\rangle=\int F([x],t)|[x]\rangle [dx]$, the result of acting 
the interaction operator is determined 
by the operation of the kernel function $V(x_3,x_4;x_1,x_2)$ 
as a two-body integral operator 
to all the possible pairs of two coordinates $(x_a,x_b)$ $(a\ne b)$ 
contained in  the 
coefficient function $F([x],t)=F(x_1,\ldots,x_N,t)$, in the following 
form:
\begin{align}
	\sum_{a,b \,(a\ne b)}^N\iint V(x_3',x_4';x_a,x_b)F(x_1,\ldots,x_N,t)d^3x_ad^3x_b ,
	\nonumber
\end{align}
with the coordinates other than the pair $(x_a,x_b)$ 
passing through freely in each term of the summation.
Consequently, the highest 
eigenvalue zero corresponding to 
equilibrium states, 
satisfying 
\begin{align}
\lim_{t\rightarrow \infty}{\hat {\mathcal H}}_I|F(t)\rangle=0, 	
\label{equicond}
\end{align}
will be realized if and only if  
$F([x],\infty)$ depends on the coordinates $x_a$ 
$(a=1,\ldots, N)$ exclusively {\it only 
through the total energies}, 
$\displaystyle\sum_{a=1}^N K(x_a)$ and $
\displaystyle\sum_{a=1}^N H(x_a)$, since 
it requires that the coordinate dependence 
of the function $F([x],t)$ is allowed only 
through $H(x_a)+H(x_b)$ and $K(x_a)+K(x_b)$ 
for all possible pairs $(x_a,x_b)$. 
Such states are not unique because any distribution function of two total 
energy functions of the 
form
\[
f\big(\displaystyle\sum_{a=1}^N K(x_a),\displaystyle\sum_{a=1}^N H(x_a)\big)\]
 is allowed. Therefore the equilibrium states as 
 eigenstates with the eigenvalue $0$ of the interaction operator ${\hat {\mathcal H}}_I$ are infinitely 
 degenerate. Obviously, the general solution of the equilibrium 
condition \eqref{equicond}, which is itself linear, is 
a linear combination of all such possible statistical states. 
Needless to say, the trivial case of 
the constant coefficient function corresponds to the $Z$-vacuum 
\eqref{consproZ}. 

Then, the next possible eigenvalue of the interaction operator 
${\mathcal H}_I$ different from zero 
is given by
\begin{align}
\label{excite}
	-\frac{g^2}{4}\big(N(N-1)-(N-1)(N-2)\big)=-\frac{1}{2}g^2(N-1),
\end{align}
which is also infinitely degenerate. This is because of the 
following arguments:
the corresponding eigenfunctions are, in general, given as linear combinations of distribution functions, 
whose dependence on the energy functions in each of them is through a pair of 
energy functions $\big(K(x_a),H(x_a)\big)$ with some $a$, and a pair of the 
sums of the remaining $N-1$ energy functions with indices other than $a$. This leads to 
the fact that for excited states we can have, {\it at most}, $2
\begin{pmatrix}
	N-1 \\ 2
\end{pmatrix}=(N-1)(N-2)$ pairs of two-body states of 
zero eigenvalue, 
instead of $2\begin{pmatrix}
	N \\ 2
\end{pmatrix}=N(N-1)$ such pairs for the ground states, 
and the difference of the number of the 
zero-eigenvalue pairs contribute to the 
final eigenvalue of the first excited states, 
yielding \eqref{excite}. 

As a consequence of \eqref{excite}, the approach to an  
equilibrium state follows in general an exponential law, 
$
e^{-g^2(N-1)t/2}, 
$
for sufficiently large $t$ for any $N\ge 2$. 
Our intuitive picture for the 
approach to equilibrium states explained in the Introduction 
is valid even for the simplest case $N=2$. Furthermore, 
no matter how $g^2$ is small, 
the relaxation time is finite for sufficient large $N\gtrsim 1/g^2$. 
 It seems appropriate to 
 say that our non-local `stochastic' interaction is 
indeed sufficiently `chaotic'.
 


\subsection{Equilibrium statistical states from the H-theorem}

Specification of equilibrium statistical states more 
refined than \eqref{equicond} is attained only when specific conditions for the initial statistical states are 
given.  
In fact, if we set up initial conditions appropriately, 
we can employ 
the minimum principle derived from the H-theorem 
to obtain the information on the 
equilibrium states under given initial conditions: 
denoting an equilibrium distribution function by 
$F_{\infty}[x]\equiv \langle [x]|F(\infty)\rangle$, the H-function 
at the equilibrium,
\begin{align}
	{\mathsf H}=\int F_{\infty}[x]\log F_{\infty}[x] [dx],
	\nonumber
\end{align} 
must take the minimum value after the evolution 
described by the master equation with the initial state function $\langle [x]|F(0)\rangle$. Therefore, 
we can apply variational arguments for deriving 
a particular set of equilibrium statistical states with 
suitable constraints arising from the initial conditions 
imposed on $\langle [x]|F(0)\rangle$. 

\subsubsection{Generalized micro-canonical states}
The simplest and meaningful initial condition conceivable is that 
both of the total values of energy functions $\displaystyle\sum_{a=1}^N K(x_a)$ and $
\displaystyle\sum_{a=1}^N H(x_a)$ have fixed numerical values
$(h_0,k_0)$:
\begin{align}
	\sum_{a=1}^N H(x_a)=h_0, \qquad \sum_{a=1}^N K(x_a)=k_0.\nonumber
\end{align} 
Since the total energies are strictly conserved during 
evolution to equilibrium states, 
we can consistently 
impose the following conditions locally with respect 
to the coordinates for the equilibrium distribution 
function,
\begin{align}
	F_{\infty}[x]\bigg(\sum_{a=1}^N H(x_a)-h_0\bigg)=0, \quad 
	F_{\infty}[x]\bigg(\sum_{a=1}^N K(x_a)-k_0\bigg)=0.\label{microca} 
\end{align}
The variational equation is then 
\begin{align}
	&\delta_{F,\alpha,\gamma_H,\gamma_K}\bigg[{\mathsf H}+\alpha\bigg(\int F_{\infty}[x][dx]-1\bigg)\nonumber\\
	&+\int \gamma_H[x]F_{\infty}[x]\bigg(
	\sum_{a=1}^N H(x_a)-h_0	\bigg)[dx]+\int \gamma_K[x]F_{\infty}[x]\bigg(
	\sum_{a=1}^N K(x_a)-h_0	\bigg)[dx]\bigg]=0.\nonumber
\end{align}
Note that in addition to the Lagrange multiplier $\alpha$
for the constraint $
\int F_{\infty}[x][dx]=1$, we have to 
introduce the Lagrange `{\it multiplier functions}' 
$\gamma_H[x]$ and $\gamma_K[x]$, corresponding to \eqref{microca}. This implies that $F_{\infty}[x]$ can have nonzero constant values only 
on the level surfaces defined by \eqref{microca}. 
The variational equation for $\delta F_{\infty}[x]$ is, after making a shift $\alpha\rightarrow \alpha-1$, 
\begin{align}
	\log F_{\infty}[x] +\alpha +\gamma_H[x]\bigg(\sum_{a=1}^N H(x_a)-h_0\bigg)
	+\gamma_K[x]\bigg(\sum_{a=1}^N K(x_a)-k_0\bigg)
	=0\nonumber
\end{align}
which must be solved together with the constraints \eqref{microca} 
obtained by the variations with respect to $(\gamma_H[x],\gamma_K[x])$. 
Let us set
\begin{align}
	F_{\infty}[x]=e^{-\alpha+f[x]}\delta\bigg(\sum_{a=1}^N H(x_a)-h_0\bigg)
	\delta\bigg(\sum_{a=1}^N 
	K(x_a)-k_0\bigg)
	\label{preF}.
\end{align}
This yields the condition for determining $f[x]$
\begin{align}
	&f[x]+\log \bigg[\delta\bigg(\sum_{a=1}^N H(x_a)-h_0\bigg)
	\delta\bigg(\sum_{a=1}^N 
	K(x_a)-k_0\bigg)\bigg]\nonumber \\
	 &+	\gamma_H[x]\bigg(\sum_{a=1}^N H(x_a)-h_0\bigg)
	+\gamma_K[x]\bigg(\sum_{a=1}^N K(x_a)-k_0\bigg)=0.
	\nonumber
\end{align}
However, the function $f[x]$ only occurs in \eqref{preF} 
with the product of the $\delta$-functions constraining 
the values of total energy functions, 
the second line can be set to zero, and 
also the logarithmic term in the 
first line gives an infinite constant: $
f[x]=-2\log \delta(0)
$. This infinite contribution is absorbed by making an infinite  
renormalization of $\alpha$: $\alpha=\alpha_0-2\log \delta(0)$.
We then arrive at 
\begin{align}\label{microcano}
	F_{\infty}[x]=e^{-\alpha_0(h_0,k_0)}\delta\bigg(\sum_{a=1}^N H(x_a)-h_0\bigg)
	\delta\bigg(\sum_{a=1}^N 
	K(x_a)-k_0\bigg)\equiv F_{{\rm micro}}[x]
	\end{align}
where 
\begin{align}
	e^{\alpha_0(h_0,k_0)}=
	\iint \cdots\int  \delta\bigg(\sum_{a=1}^N H(x_a)-h_0\bigg)
	\delta\bigg(\sum_{a=1}^N 
	K(x_a)-k_0\bigg)[dx].\nonumber
\end{align}
Notice that this result satisfies the condition 
that all of the phase-space coordinates appear only 
through the total energy functions and hence gives a special solution 
of the equilibrium condition \eqref{equicond} as it should.

\subsubsection{Generalized canonical states}

As an alternative to the condition \eqref{microca}, we can 
require a weaker condition, namely, that the expectation values of the total energies are fixed to be a set of 
numerical values $(k,h)$ as the initial condition.
In fact, the expectation values, $\langle Z|\hat{H}|F(t)\rangle$ and 
$\langle Z|\hat{K}|F(t)\rangle$, of total energy functions are also 
guaranteed to be conserved under the evolution governed by 
the master equation:
\begin{align}
\begin{aligned}
	&\frac{d}{dt}\langle Z|\hat{H}|F(t)\rangle
	=\langle Z|\hat{H}{\hat {\cal H}}_I(t)|F(t)\rangle
	=\langle Z|[\hat{H},{\hat {\cal H}}_I(t)]|F(t)\rangle=0,\\
	&\frac{d}{dt}\langle Z|\hat{K}|F(t)\rangle
	=\langle Z|\hat{K}{\hat {\cal H}}_I(t)|F(t)\rangle
	=\langle Z|[\hat{K},{\hat {\cal H}}_I(t)]|F(t)\rangle=0,
\end{aligned}
\label{consmeanenergy}
\end{align}
due to \eqref{commuHKint} and \eqref{consproZ}. 
Therefore we can set up the variational equation consistently for the 
equilibrium distribution function $F_{\infty}[x]$ as
\begin{align}
	&\delta_{F,\alpha,\beta_H,\beta_K}\bigg[\mathsf{H}+\alpha\int F_{\infty}[x][dx]\nonumber \\
	&+
	\beta_h\bigg(\int F_{\infty}[x]\sum_{a=1}^NH(x_a)[dx]-h\bigg)
	+\beta_k\bigg(\int F_{\infty}[x]\sum_{a=1}^NK(x_a)[dx]-k\bigg)
	\bigg]=0,\nonumber
\end{align}
where we introduced the Lagrange multiplier 
$\beta_h$ and $\beta_k$ corresponding to the 
initial conditions stated above:
\begin{align}
\label{hkconst}
	\int F_{\infty}[x]\sum_{a=1}^NH(x_a)[dx]=h, \quad \int F_{\infty}[x]\sum_{a=1}^NK(x_a)[dx]=k.
\end{align}
We immediately obtain, after a redefinition $\alpha\rightarrow 
\alpha-1$,
	\[
	\log F_{\infty}[x]=-\alpha -\sum_{a=1}^N\big[\beta_hH(x_a)+\beta_kK(x_a)
	\big]
	\]
which gives
\begin{align}
	F_{\infty}[x]&=\exp\bigg\{-\alpha -\beta_h\sum_{a=1}^NH(x_a)-\beta_k\sum_{a=1}^NK(x_a)
\bigg\}\equiv F_{(\beta_h,\beta_k)}[x], \label{genecano}\\
	e^{\alpha}&=\int \exp\bigg\{-\beta_h\sum_{a=1}^NH(x_a)-\beta_k\sum_{a=1}^NK(x_a)\bigg\}[dx],\nonumber\\
	h&=\int \sum_{a=1}^NH(x_a)
	\exp\bigg\{-\alpha-\beta_h\sum_{a=1}^NH(x_a)-\beta_k\sum_{a=1}^NK(x_a)\bigg\}[dx],\nonumber\\
	k&=\int \sum_{a=1}^NK(x_a)
	\exp\bigg\{-\alpha-\beta_h\sum_{a=1}^NH(x_a)-\beta_k\sum_{a=1}^NK(x_a)\bigg\}[dx],\nonumber
\end{align}
where the last three equations implicitly determine 
the value of three Lagrange multiplies $(\alpha,\beta_h,\beta_k)$. 
The free energy 
of this system is $-\alpha$, and the entropy is 
$-{\mathsf H}$. We call the distribution function \eqref{genecano} 
equipped with two independent temperatures $(1/\beta_h, 
1/\beta_k)$, 
the `generalized canonical distribution', that extends 
the ordinary canonical distribution with a 
single temperature to that with a set of two temperatures. Of course, this is what Nambu 
expected in his original paper, but 
is now justified from a genuinely dynamical standpoint on the basis of our foregoing discussions 
including the crucial role of the non-local interaction, 
without relying on probabilistic argument.

Again, this result satisfies the requirement that 
the phase-space coordinates occur only through 
the total energies and hence gives another special solution 
to \eqref{equicond}. Furthermore, since the total 
energies are simply sums of the contribution 
from each set of coordinates of $N$ constituent 
systems of 
the ensemble, the distribution function 
takes a factorized form: 
\begin{align}
	F_{(\beta_h,\beta_k)}[x]=\prod_{a=1}^NF_{(\beta_h,\beta_k)}(x_a), \quad F_{(\beta_h,\beta_k)}(x)\equiv e^{-\bar{\alpha}-\beta_hH(x)-
	\beta_kK(x)},\nonumber
\end{align}
with the normalization condition $\int F_{(\beta_h,\beta_k)}(x)d^3x=1$,
and the total energies are given as
\begin{align}
	&h\equiv N\bar{h}, \quad \bar{h}=e^{-\bar{\alpha}}\int 
	H(x)e^{-\beta_hH(x)-
	\beta_kK(x)}d^3x, \nonumber\\ 
	&  
	k\equiv N\bar{k}, \quad \bar{k}=e^{-\bar{\alpha}}\int K(x)e^{-\beta_hH(x)-
	\beta_kK(x)},\nonumber
	\\
	&e^{\bar{\alpha}}=\int e^{-\beta_hH(x)-
	\beta_kK(x)}d^3x=e^{\alpha/N}.\nonumber
\end{align}
Thus the corresponding `generalized canonical state' is 
simply expressed as
\begin{align}
	&|F(\infty)\rangle_{(\beta_h,\beta_k)}\equiv \frac{1}{\sqrt{N!}}\Psi^{\dagger
}(\beta_h,\beta_k)^N|0\rangle, \label{genecanostate}\\
	&\Psi^{\dagger}(\beta_h,\beta_k)=\frac{\int e^{-\beta_hH(x)-\beta_kK(x)}
	\psi^{\dagger}(x,t_0)d^3x}{\int e^{-\beta_hH(x)-
	\beta_kK(x)}d^3x},\nonumber\\ \,\,&\Psi(\beta_h,\beta_k)=\frac{\int e^{-\beta_hH(x)-\beta_kK(x)}
	\psi(x,t_0)d^3x}{\int e^{-\beta_hH(x)-
	\beta_kK(x)}d^3x},\nonumber
\end{align}
with the normalization condition
\begin{align}
	\langle Z|F(\infty)\rangle_{(\beta_h,\beta_k)}=1.
	\nonumber
\end{align}
The operators $\Psi^{\dagger}(\beta_h,\beta_k), 
\Psi(\beta_h,\beta_k)$, which are to be called 
`thermal field operators' with a set of inverse temperatures 
$(\beta_h,\beta_k)$, are independent of 
an arbitrary time parameter $t_0$ 
residing in their integral representations, owing to 
the field equations \eqref{ffeq} for $(\psi,\psi^{\dagger})$, 
and are characterized by the commutation relations:
\begin{align}
	[\psi(x,t_0), \Psi^{\dagger}(\beta_h,\beta_k)]
	=e^{-\alpha_0-\beta_hH(x)-\beta_kK(x)}
	=[\Psi(\beta_h,\beta_k), \psi^{\dagger}(x,t_0)].\nonumber
\end{align}

\subsubsection{Statistical systems combined with different temperatures as an initial state}
Notice, as a matter of course, that the states generated by the thermal field 
operators can be equilibrium states if and only 
if the set of temperatures of the constituent 
systems of the ensemble are all the same. As a simple example of 
non-equilibrium initial states characterized by non-uniform temperatures, 
let us consider an initial statistical state constructed 
by a product of the thermal field operators 
with different sets of temperatures such as, say,  
\begin{align}
	|F(0)\rangle_{(1)(2)}\equiv \frac{1}{N_{(1)(2)}}
	\Psi^{\dagger}(\beta^{(1)}_k,\beta^{(1)}_h)^{N_1}
	\Psi^{\dagger}(\beta^{(2)}_k,\beta^{(2)}_h)^{N_2}
	|0\rangle, \nonumber
\end{align}
with the normalization constant $N_{(1)(2)}$ being fixed by $\langle Z|F(0)\rangle_{(1)(2)}=1$. 
Then, after an infinite evolution, the corresponding equilibrium statistical state 
will be obtained in the form \eqref{genecanostate}: this is 
guaranteed due to the minimum principle used to 
derive the generalized canonical statistical state. It then 
follows that we have the following expansion,
\begin{align}
	|F(0)\rangle_{(1)(2)}
	=|F(\infty)\rangle_{(\beta_h,\beta_k)}
	+\sum_{L}c_L|F_{L<0}\rangle, \nonumber
\end{align}
where the new set of inverse temperatures $(\beta_h,\beta_k)$ 
in the equilibrium is determined by the 
expectation value of total energies 
$\langle Z|\hat{H}|F(0)\rangle_{(1)(2)}, \langle Z|\hat{K}|F(0)\rangle_{(1)(2)}$
of this 
initial statistical state, and, in the second term, $|F_{L<0}\rangle$ 
are the $N$-body eigenstates 
of the interaction operator $\hat{{\mathcal H}}_I$ 
with {\it negative} eigenvalues $L$, which by definition satisfy 
the orthogonality condition $\langle Z|F_{L<0}\rangle=0$. 
During evolution to $t=\infty$, the latter terms are 
fading away exponentially; in general, for any 
operator $\hat{O}$ which is commutative with $\hat{{\mathcal H}}_I$, the condition \eqref{consproZ} 
shows that, because of the arbitrariness of $t$,
\begin{align}
	\langle Z|\hat{O}|F_{L<0}\rangle=\langle Z|e^{t\hat{{\mathcal H}}_I}\hat{O}|F_{L<0}\rangle=\langle Z|\hat{O}e^{t\hat{{\mathcal H}}_I}|F_{L<0}\rangle=\lim_{t\rightarrow \infty}\langle Z|\hat{O}e^{t\hat{{\mathcal H}}_I}|F_{L<0}\rangle=0. 
	\nonumber
\end{align}
This ensures the equalities
\begin{align}
	&\langle Z|\hat{H}|F(0)\rangle_{(1)(2)}=
	\langle Z|\hat{H}|F(\infty)\rangle_{(\beta_h,\beta_k)}, 
	\quad \langle Z|\hat{K}|F(0)\rangle_{(1)(2)}=
	\langle Z|\hat{K}|F(\infty)\rangle_{(\beta_h,\beta_k)}.\nonumber
\end{align}

\vspace{0.2cm}
\subsection{The generalized KMS-like conditions }

Let us consider, in the case of a 
generalized canonical distribution, the expectation value of an arbitrary `sum function' $O([x],t)$ 
of the form
\begin{align}
	O([x],t)=\sum_{a=1}^NO(x_a,t),\nonumber
\end{align}
corresponding to the operator $\hat{O}(t)=\int 
\psi^{\dagger}(x,t)O(x)\psi(x,t)d^3x$. Then, 
it is decomposed into a sum of the expectation values for 
each single Nambu system, and hence it is sufficient to 
consider
\begin{align}
	\langle O(x,t)\rangle_{(\beta_h,\beta_k)}	\equiv e^{-\bar{\alpha}}
	\int O(x,t)e^{-\beta_hH(x)-\beta_kK(x)}d^3x.\nonumber
\end{align}

If we choose a special case involving a Nambu bracket in 
the form of the Nambu equations of motion as 
$
	O=B\{H,K,A\}\equiv -B\{K,H,A\}\equiv B\dot{A}
$ with two functions $A=A(x,t)$ and $B=B(x,t)$ 
we have the 
following conditions:
\begin{align}
	\beta_h\langle B\dot{A}\rangle_{(\beta_h,\beta_k)} =\langle
	\{K,A,B\}\rangle , 
	\quad -\beta_k\langle B\dot{A}\rangle_{(\beta_h,\beta_k)} =\langle
	\{H,A,B\}\rangle=\langle \{A,H,B\}\rangle,\label{KMSlike}
\end{align}
as immediate consequences of the identities, 
\begin{align}
	0=\int \{K,A,B\,e^{-\beta_hH-\beta_kK}\}d^3x,\quad 0=\int \{H,A,B\,e^{-\beta_hH-\beta_kK}\}d^3x,\nonumber
\end{align}
respectively.  Note that these relations are the special cases of a 
more general identity 
with three functions $(A,B,C)$: 
\begin{align}
	0&=\int  \{A,B,Ce^{-\beta_hH-\beta_kK}\}d^{3}x\nonumber 
	\\&=
\int \{A,B,C\}e^{-\beta_hH-\beta_kK}d^3x
-\int C\{A,B,\beta_hH+\beta_kK\}e^{-\beta_hH-\beta_kK}d^3x.
\nonumber
\end{align}
Here, the integrand must be assumed to be 
well-behaved at asymptotic infinities $|x|\rightarrow \infty$ 
to ensure the vanishing of total derivative 
involved in this formula. 

In fact, the above relations are straightforward generalizations of 
that known in conventional Hamiltonian dynamics 
for the Poisson bracket $\{A,B\}$ and a single Hamiltonian $H$:
\begin{align}
	\beta \langle B\dot{A}\rangle \equiv \beta\langle B\{A,H\}\rangle_{\beta}=\langle \{A,B\}\rangle, \label{mermin}
\end{align}
where 
$
	\langle O\rangle_{\beta}\equiv \int Oe^{-\beta H} d\Gamma /\int 
	e^{-\beta H}d\Gamma
$
for any observable $O$ with $d\Gamma$ being the volume element of the phase space. 
The relevant identity is 
\begin{align}
	0=\int  \{A, Be^{-\beta H}\}d\Gamma=\int \{A,B\}e^{-\beta H}
	d\Gamma-
\beta\int B\{A,H\}e^{-\beta H}d\Gamma.\nonumber
\end{align}
The relation of this type has been 
utilized successfully in, e.g., reference\cite{mermin} in analyzing various 
models in classical statistical mechanics. 
Furthermore, because 
\eqref{mermin} connects the time derivative $\beta\dot{A}$ on the left-hand side to the Poisson bracket operation $\{A,B\}$ 
on the right-hand side, 
it can be related, in the classical limit\cite{galla} to the well-known 
KMS condition  (Kubo, Martin, Schwinger; for a precise and self-contained 
account with extensive bibliography, see the reference\cite{haag}) 
which plays important roles in characterizing the 
equilibrium states in the standard {\it quantum} (equilibrium) statistical mechanics by the method of analytic continuation of time variable in the complex plane of time. 
Indeed, the KMS condition using a finite (imaginary) shift of time 
can be expressed in the form
$$\omega_{\beta}( [A(t),B(t)]/i\hbar) =
\omega_{\beta}(B(t)\big(A(t+i\hbar \beta)-A(t)\big)/i\hbar)\rangle)$$ 
where $\omega_{\beta}(\cdots)$ denotes the quantum mechanical 
thermal expectation value in terms of the density 
matrix $\rho_{\beta}$: $\omega_{\beta}={\rm Tr}(\rho_\beta \cdots)$. The imaginary 
unit is cancelled when we take the classical 
limit $\hbar\rightarrow 0$ by replacing the commutator with the Poisson bracket, 
$[A,B]/i\hbar \rightarrow \{A,B\}$ and $\big(A(t+i\hbar \beta)-A(t)\big)/i\hbar\rightarrow \beta\dot{A}$. 

In this way, \eqref{KMSlike} in Nambu dynamics has a close analogy with the conventional Hamiltonian dynamics replacing the Poisson bracket with Nambu bracket and the single Hamiltonian $H$ 
in the former either with $K$ or $H$ in the latter: therefore it seems appropriate to call \eqref{KMSlike} the `generalized KMS-like conditions'. 
The conditions \eqref{KMSlike} which 
relate the doublet $(\beta_h\dot{A},\beta_k\dot{A})$ to that of 
the Nambu brackets $(\{K,A,B\}, \{A,H,B\})$, might be quite 
suggestive in attempting quantization of Nambu dynamics.
In fact, the analogy between the role of a Nambu bracket and that of a Poisson 
bracket in the expectation values 
becomes more acute 
if we recall that, as is well known, the Nambu 
bracket with fixed $K$ or $H$, denoted respectively as
\begin{align}
	\{A,B\}_K\equiv \{K,A,B\} \quad {\rm or} \quad 
	\{A,B\}_H\equiv \{A,H,B\}=-\{H,A,B\},
\end{align}
satisfying $\{A,K\}_K=0$ or $\{A,H\}_H=0$ for 
any function $A$, can be interpreted as a generalized Poisson bracket,  
which is compatible with a constraint $K=$constant or $H=$constant,  
satisfing the Jacobi identity.
These properties ensure that in the Nambu equations of motion 
the generalized Poisson 
bracket $\{A,B\}_K$ or $\{A,B,\}_H$ plays the role 
of the ordinary Poisson bracket depending on 
whether $H$ or $K$, respectively, is treated as the Hamiltonian for 
time evolution (see, e.g., reference \cite{yoneintro}). This indeed suggests  
a doublet structure in the form \eqref{KMSlike} for interpreting 
time evolution in the Nambu dynamics.  A related viewpoint has been discussed in \cite{yoneHJ} (which the interested readers are referred to including 
relevant literature) from the standpoint of a generalized Hamilton-Jacobi formulation 
and its application to a possible Schr\"{o}dinger-type (wave-mechanical) quantization of 
Nambu dynamics.

\subsection{The `relativity' of temperatures -- the $\mathcal{N}$ symmetry of the equilibrium 
states}
We have stressed the  importance of the $\mathcal{N}$-symmetry with the SL(2,${\mathbb R}$) group in our development of interacting Nambu dynamics. 
Let us finally consider the meaning, if any, of this symmetry 
for the equilibrium statistical states. 

We first notice that both of the generalized microcanonical state 
and generalized canonical state exhibit the corresponding 
symmetry, provided that we assign transformation laws for 
$(h_0,k_0)$ and $(\beta_h,\beta_k)$ appropriately: 
\begin{enumerate}
\item[(A)] The microcanonical distribution \eqref{microcano} is invariant under \eqref{sl2r}, 
if 
\begin{align}
	\begin{pmatrix}
		h_0' \\k_0'
	\end{pmatrix}=\begin{pmatrix}
		 a & b \\ c & d 
	\end{pmatrix}\begin{pmatrix}
		h_0 \\k_0
	\end{pmatrix}.\nonumber
\end{align}
This is clear from the equality
\[
\delta\bigg(\sum_{a=1}^N H(x_a)-h_0\bigg)
	\delta\bigg(\sum_{a=1}^N 
	K(x_a)-k_0\bigg)=\delta\bigg(\sum_{a=1}^N H'(x_a)-h_0'\bigg)
	\delta\bigg(\sum_{a=1}^N 
	K'(x_a)-k_0'\bigg).\]

\item[(B)] The generalized canonical distribution 
\eqref{genecano} is invariant, if 
\begin{align}
	\begin{pmatrix}
		\beta_h' & \beta_k'
	\end{pmatrix}
	=\begin{pmatrix}
		\beta_h & \beta_k
	\end{pmatrix}	\begin{pmatrix}
		d & -b \\
		-c & a
	\end{pmatrix}, \nonumber 
\end{align}
or, equivalently, $(-\beta_k,\beta_h)$ transforms in the same way as  
$(H,K)$, 
such that the bilinear form appearing in the statistical weight 
satisfies
\begin{align}
	\beta_h'H'(x)+\beta_k'K'(x)=\beta_hH(x)+\beta_kK(x).
	\nonumber
\end{align}
\end{enumerate}
Also, it is to be noticed that the vector-like generalized KMS-like condition  \eqref{KMSlike},  
consisting of two components, are consistent with 
the ${\mathcal N}$ symmetry in the sense that they are covariant under the ${\mathcal N}$ transformations. 
These symmetries are meaningful only when 
we consider the transformations in the spaces of all possible values of energies $(h_0,k_0)$ or 
inverse temperatures $(\beta_h,\beta_k)$, respectively, which can be 
regarded as being {\it dual} to each other. A single system 
with particular numerical values of the energies in (A) and 
inverse temperatures in (B), the ${\mathcal N}$ symmetry is, so to speak, broken spontaneously. 


Let us here recall that the descriptions 
of interacting Nambu dynamics in terms of the energy functions $(H,K)$ or 
$(H',K')$ can be regarded as completely equivalent to each other 
and hence should not be discriminated, 
since the equations of motion and the interactions are 
invariant under the SL(2,${\mathbb R}$) transformations 
(as the analogue to Lorentz transformations). 
This implies that a set of inverse temperatures has no `absolute' significance 
as a characterization of the generalized canonical states, 
but merely has a `relative' significance: the temperatures are defined 
only after a pair of energy functions is chosen among a 
continuously possible equivalent class of energy functions
(in analogy with the choice of a particular inertial frame in 
relativity theory), connected by the SL(2,${\mathbb R}$) transformations \eqref{sl2r} under which 
the equilibrium statistical states are covariant.  

In particular, 
since the number (two) of the degrees of freedom of the set of inverse temperatures are surpassed by the number (three) of 
the degree of SL(2,${\mathbb R}$) group, 
the two-dimensional inverse-temperature plane $(\beta_h,\beta_k)$ 
can essentially be covered starting 
with an initial point, say, $(0,1)$, by the SL(2,${\mathbb R}$) transformations:  
for instance, if we employ a well-known standard parametrization 
(called 
the `Iwasawa decomposition') of 
an arbitrary element of SL(2,${\mathbb R}$),
\[
g=\begin{pmatrix}
	1 & x \\ 0 &1
\end{pmatrix}\begin{pmatrix}
	e^{s} & 0 \\ 0 &e^{-s}
\end{pmatrix}
\begin{pmatrix}
	\cos \theta & -\sin\theta \\ \sin\theta & \cos\theta
\end{pmatrix}, \quad (x,s)\in {\mathbb R}^2, \quad
0\le \theta \le 2\pi, 
\]
we can set
\begin{align}
	(\beta_h,\beta_k)=(0,1)g=(0,e^{-s})
\begin{pmatrix}
	\cos \theta & -\sin\theta \\ \sin\theta & \cos\theta
\end{pmatrix}=e^{-s}(\sin\theta, \cos\theta),\nonumber
\end{align}
in which the whole of the first quadrant 
($0\le  \theta \le \pi/2$) of 
(positive) inverse-temperature plane is covered with the two 
parameters $(s,\theta)$, except for the 
origin and infinities corresponding to the singular limit 
$|s|\rightarrow \infty$. 

By contrast, the ordinary temperature $T=1/\beta$, characterizing an 
ordinary canonical 
ensemble with Boltzmann factor $e^{-\beta H}$,  
has an absolute meaning once the Hamilton 
equations of motion are given. From a conceptual 
viewpoint of statistical physics, this is an important and critical difference 
by which the Nambu dynamics departs 
from the conventional Hamiltonian dynamics. 




%

\vspace{1cm}
\appendix 
\setcounter{equation}{0}
\noindent
\begin{center}
	{\Large Appendix}
\end{center}
 
\section{Stochastic Nambu equations of motion}
\renewcommand{\theequation}{A.\arabic{equation}}
In the present paper, we have developed a statistical field theory 
of many-body Nambu dynamics in which an approach to 
statistical equilibrium is attained through an 
autonomous Markov process caused by a non-local self-interaction   
among Nambu particles. 
We can also consider a more phenomenological approach 
in the sense that we focus only on a 
single Nambu particle, regarding effectively all the other Nambu particles as a whole 
to be the environment (or heat bath) with a definite 
set of temperatures, provided that the 
interactions are sufficiently weak to make such a 
picture feasible. 

Let us start with the Langevin equation in $n$ dimensions,
\begin{align}
	\frac{dx^i}{dt}=f^i(x)+r^i(t)\nonumber
\end{align}
where $f^i(x)$ is the time independent vector force-field and $r^i(t)$ is a random noise that is treated as 
a stochastic variable with mean values:
\begin{align}
	\langle r^i(t)\rangle=0, \quad 
	\langle r^i(t)r^j(t')\rangle=2D\delta^{ij}(t-t'),
	\nonumber
\end{align}
$D$ being the diffusion constant. As is well-known, 
the probability density $P(x,t)$ for the 
coordinates $x^i$ satisfies the equation of continuity,
\begin{align}
	\frac{\partial P(x,t)}{\partial t}+\frac{\partial J^i(x,t)}{\partial x^i}=0, 
	\quad J^i(x,t)=\bigg(f^i(x) -D\frac{\partial}{\partial x^i}
	\bigg)P(x,t),\nonumber
\end{align}
the so-called Fokker-Planck equation (see, e.g., the reference \cite{radpay} for a recent review). 

It follows in general that the 
distribution function $P(x,\infty)$ at an equilibrium 
satisfy
\begin{align}
	\frac{\partial}{\partial x^i}\bigg[\bigg(f^i(x) -D\frac{\partial}{\partial x^i}
	\bigg)P(x,t)\bigg]=0.\nonumber
\end{align}
If we assume a generalized canonical distribution of 
Nambu dynamics in $n$-dimensions with 
$n-1$ energy functions $H_k$ $(k=1,\ldots,n-1)$ 
and the corresponding temperatures $\beta_k$ as 
\begin{align}
	P(x,\infty)=e^{-\bar{\alpha}-H_{\beta}(x)}, 
	\quad H_{\beta}(x)\equiv \sum_{k=1}^{n-1}\beta_kH_k(x),
	\nonumber
\end{align}
we have a condition for the vector force-field $f^i$:
\begin{align}
	\partial_if^i-f^i\partial_iH_{\beta}
	+D\partial_i^2H_\beta-D(\partial_iH_{\beta})^2=0.\nonumber
\end{align}
The solution of this equation which reduces to the Nambu equations of motion 
in the limit $D\rightarrow 0$ is 
\begin{align}
	f^i=X^i-D\partial_iH_{\beta}, \quad 
	X^i=\epsilon^{ij_1\cdots j_{n-1}}
	\partial_{j_1}H_1\cdots\partial_{j_{n-1}}H_{n-1}.\nonumber
\end{align}

This implies, conversely, that there exists a class 
of initial distribution functions $P(x,0)$ for which the  
stochastic differential equations of motion
\begin{align}
	\frac{dx^i}{dt}=X^i-D\partial_iH_{\beta}+r^i(t), \nonumber
\end{align}
yields the generalized 
canonical distribution for $t\rightarrow \infty$. Thus the second 
and third terms on the right-hand side of this equation can 
be interpreted as representing the environmental force. 
In particular, the second term is a dissipating frictional force 
associated with the presence of random fluctuations caused by $r^i(t)$. 

Note that, since we are here treating the case of $n-1$ energy functions, the group of 
the ${\mathcal N}$-symmetry is SL($n-1,\mathbb{R}$), and, 
correspondingly, the transformations of the set of temperatures $(\beta_1,\ldots,\beta_{n-1})$ obey the fundamental 
vector representation of SL($n-1,\mathbb{R}$). 

To derive this kind of effective 
descriptions directly from the formalism of the main text 
by making a concrete separation of dynamical variables 
between a single Nambu particle and others as a thermal 
environment would be an interesting challenge. 

\section{An analog-matrix model for the kernel function 
$v(x_3,x_4;x_1,x_2)$}
\renewcommand{\theequation}{B.\arabic{equation}}

Consider an $n\times n$ `positive' matrix $p_{ij}^{(n,n)}$ 
with entries
$
p_{ij}^{(n,n)}=p_{ji}^{(n)}=1/n, 
$
satisfying
\begin{align}
	\sum_k p_{ik}^{(n,n)}p_{kj}^{(n,n)}=p_{ij}^{(n,n)}, 
	\quad
	\sum_ip_{ij}^{(n,n)}=1=\sum_jp_{ij}^{(n,n)}, 
	\label{doublestochas}
\end{align}
analogously to \eqref{normalcond2} and 
\eqref{iiiv1}.
It is easy to confirm directly that only possible eigenvalues 
of the matrix $p_{ij}^{(n,n)}$ are 1 and 0, where 
the eigenvalue 1 is not degenerate, but 
the eigenvalue 0 is degenerate with the degeneracy $(n-1)$.  
In fact, the eigenvector $r_i^{(n)}$ corresponding to the eigenvalue 1 is 
$r_i^{(n)}=1$ for all $i=1,\ldots, n$, as an analogy to 
\eqref{constf}:
\[
\sum_jp_{ij}^{(n,n)}r_j=n\times 1/n=1.
\]
The eigenvectors corresponding to the eigenvalue 0 
constitute $n-1$ dimensional vector space 
consisting of arbitrary vectors $s_i^{(n)}$ such that 
$
\sum_is_i^{(n)}=0,
$
as an analogy to \eqref{firstexcite}.

Furthermore, if we do not set any bound for $n$, we are 
naturally led to consider an infinite-dimensional `non-negative' symmetric matrix $P$
which consists of an infinite number of block diagonal 
matrices $p^{(n,n)}$ of all $n\ge 2$: 
\begin{align}
	P=\begin{pmatrix}
		p^{(2,2)} & 0^{(2,3)} & 0^{(2,4)} & \cdot & \cdot & \cdots \\
		0^{(3,2)} & p^{(3,3)} & 0^{(3,4)} & \cdot & \cdot & \cdots\\
		0^{(4,2)} & 0^{(4,3)} & p^{(4,4)} & \cdot & \cdot & \cdots\\
		. & . & . &. &. &\cdots \\
		\cdot & \cdot & \cdot & \cdot & \cdot &\cdots			\end{pmatrix},\nonumber
\end{align}
satisfying 
\[
\sum_{a=1}^{\infty}P_{ab}=1=\sum_{b=1}^{\infty}P_{ab}.
\]
Here $0^{(m,n)}$ is a $m\times n$ matrix, all of
whose entries are zero. It is clear by definition itself 
that only allowed eigenvalues 
are still $1$ and $0$, but both of eigenvalues are now 
infinitely degenerate because we must take into account all 
possible (nontrivial) values for the integers $n,m$. 
Let us denote the bases for the eigenvectors  
with eigenvalue $1$ and those with eigenvalue $0$ by $R^{(n)}$ and 
$S^{(n)}$, respectively, which of course 
satisfy the orthogonality condition $\big(R^{(n)}\big)^TS^{(n)}=0$ 
in the infinite-dimensional vector space. 
They are given by 
\[
R^{(n)}=\begin{pmatrix}
	0^{(2)} \\ 0^{(3)} \\ \cdot \\ \cdot 
	\\ 0^{(n-1)} \\ r^{(n)} \\ 0^{(n+1)} \\ \cdot \\ \cdot 
	\\ \cdot \\ 
\end{pmatrix}, 
\qquad
S^{(n)}=\begin{pmatrix}
	0^{(2)} \\ 0^{(3)} \\ \cdot \\ \cdot 
	\\ 0^{(n-1)} \\ s^{(n)} \\ 0^{(n+1)} \\ \cdot \\ \cdot 
	\\ \cdot \\ 
\end{pmatrix}\]
where $r^{(n)}$ is an $n$-vector whose entries are all 1, 
while $0^{(n)}$ is an $n$-vector whose entries are all 0, 
and the set of all possible $s^{(n)}$ itself forms an $(n-1)$-dimensional vector 
subspace, as described above. 

Compared with the case of $v(x_3,x_4;x_1,x_2)$, specifying the integer $n$ corresponds 
to considering eigenvalues with fixed energies. The vanishing of the off-diagonal block matrices $0^{(m,n)}$ $(m\ne n)$
of the matrix $P$ corresponds to the energy conservations. 
The value of $n$ itself also plays the 
role of volumes of intersections of level 
energy surfaces. In the case of $v(x_3,x_4;x_1,x_2)$, the set of allowed points 
constituting a subspace 
with fixed values of energy functions becomes finite if we discretized the phase space. Remember that the space of fixed energies is compact due to our assumptions 
for the energy functions. 
Of course, the generic eigenstates consist 
of arbitrary linear combinations of these basis states 
for each eigenvalue 1 and 0. 
Therefore, eigenstates in each case form an infinite-dimensional space.


\end{document}